\documentclass{emulateapj}

\begin{document}

\newcommand{\cl}{{RX~J0152}}
\newcommand{\ms}{{MS~1054}}
\newcommand{\clt}{{CL~1358}}
\newcommand{\mst}{{MS~2053}}

\newcommand{\V}{$V_{606}$}
\newcommand{\ra}{$r_{625}$}
\newcommand{\ia}{$i_{775}$}
\newcommand{\z}{$z_{850}$}

\newcommand{\ntot}{674}
\newcommand{\nlum}{452}
\newcommand{\nmass}{441}
\newcommand{\hdf}{92}
\newcommand{\ldf}{83}

\newcommand{\etal}{{\em et~al.\,}}
\title{Mass-Selection and the Evolution of the Morphology-Density
  Relation from $z=0.8$ to 0\altaffilmark{1,2}}
\altaffiltext{1}{Based on observations with the NASA/ESA {\em Hubble Space
Telescope}, obtained at the Space Telescope Science Institute, which is
operated by the Association of Universities for Research in Astronomy,
Inc. under NASA contract NAS5-26555.}

\altaffiltext{2}{This paper includes data gathered with the 6.5 m
  Magellan Telescopes located at Las Campanas Observatory, Chile.}

\author{B. P. Holden\altaffilmark{3}}

\author{G. D. Illingworth\altaffilmark{3}}

\author{M. Franx\altaffilmark{4}}

\author{J.~P. Blakeslee\altaffilmark{5}}

\author{M. Postman\altaffilmark{6}}

\author{D.~D. Kelson\altaffilmark{7}}

\author{A. van der Wel\altaffilmark{8}}

\author{R. Demarco\altaffilmark{8}}

\author{D.~K. Magee\altaffilmark{2}}

\author{K.-V. Tran\altaffilmark{4,9}}

\author{A. Zirm\altaffilmark{8}}

\author{H. Ford\altaffilmark{8}}

\author{P. Rosati\altaffilmark{10}}

\author{N. Homeier\altaffilmark{8}}

\altaffiltext{3}{UCO/Lick Observatories, University of California,
  Santa Cruz, CA 95064; holden@ucolick.org; gdi@ucolick.org;
  magee@ucolick.org} 
\altaffiltext{4}{Leiden Observatory, Leiden
  University, P.O.Box 9513, 2300 RA, Leiden, The Netherlands;
  franx@strw.leidenuniv.nl; vy@strw.leidenuniv.nl}
\altaffiltext{5}{Department of Physics and Astronomy, Washington State
  University; Pullman, WA 99164; jblakes@wsu.edu}
\altaffiltext{6}{Space Telescope Science Institute, Baltimore, MD
  21218; postman@stsci.edu}
\altaffiltext{7}{Observatories of the
  Carnegie Institution of Washington, Pasadena, CA, 91101;
  kelson@ociw.edu}
\altaffiltext{8}{Department of Physics \&
  Astronomy, Johns Hopkins University, Baltimore, MD 21218;
  wel@pha.jhu.edu; demarco@pha.jhu.edu; azirm@pha.jhu.edu;
  ford@pha.jhu.edu;}
\altaffiltext{9}{Harvard-Smithsonian Center for
  Astrophysics, 60 Garden Street, Cambridge MA 02138}
\altaffiltext{10}{European Southern Observatory,
  Karl-Schwarzschild-Strasse 2, D-85748, Garching, Germany;
  prosati@eso.org}

\shorttitle{Mass-Selected Morphology-Density Relation}

\begin{abstract}

  We examined the morphology-density relations for galaxy samples
  selected by luminosity and by mass in each of five massive X-ray
  clusters from $z=0.023$ to $0.83$ for \ntot\
  spectroscopically-confirmed members.  Rest-frame optical colors and
  visual morphologies were obtained primarily from {\em Hubble Space
    Telescope} images. The visual morphologies ensure consistency with
  the extensive published results on galaxy evolution in dense
  environments.  Morphology-density relations (MDR) are derived in
  each cluster from a complete, luminosity-selected sample of \nlum\
  galaxies with a magnitude limit $M_V < M^{\star}_{V} + 1$.  The
  change in the early-type fraction with redshift matches previous
  work for massive clusters of galaxies.  We performed a similar
  analysis, deriving MDRs for complete, mass-selected samples of
  \nmass\ galaxies with a mass-limit of $10^{10.6} M_{\sun}$.  Our
  mass limit includes faint objects, the equivalent of $\simeq1$ mag
  below ${L^{\star}}$ for the red cluster galaxies, and encompasses
  $\simeq$70\% of the stellar mass in cluster galaxies. The MDRs in
  the mass-selected sample at densities of $\Sigma > 50\ {\rm galaxies\
    Mpc^{-2}}$ are similar to those in the luminosity-selected sample
  but show larger early-type fractions, with a weak indication of a
  shallower slope.  However, the trend with redshift in the fraction
  of elliptical and S0 galaxies with masses $> 10^{10.6} M_{\sun}$ differs
  significantly between the mass- and luminosity-selected samples.
  The clear trend seen in the early-type fraction from $z=0$ to
  $z\simeq 0.8$ is not found in mass-selected samples.  The early-type
  galaxy fraction changes much less, and is consistent with being
  constant at \hdf\% $\pm$ 4\% at $\Sigma> 500\ {\rm galaxies\ Mpc^{-2}}$
  and \ldf\% $\pm$ 3\% at $ 50\ < \Sigma < 500\ {\rm galaxies\ Mpc^{-2}}$.
  Given the mass-limit in our sample, this suggests that galaxies of
  mass lower than $> 10^{10.6} M_{\sun}$ play a significant role in
  the evolution of the early-type fraction in luminosity-selected
  samples, i.e., they are larger contributors to the
  luminosity-selected samples at higher redshifts than at low
  redshift.

\end{abstract}

\section{Introduction}

At low redshift, the cores of clusters of galaxies are dominated by
massive elliptical and S0 galaxies \citep{smith1935, zwicky1942,
  oemler1974}.  \citet{dressler1980a} quantified this as the
morphology-density relation (MDR), where the fraction of elliptical
and S0 galaxies (hereafter early-type galaxies) increases with
increasing local galaxy density.  Much discussion has centered on the
astrophysical reasons for this relation, with contributions from many
studies. For example, how galaxy morphology depends on the underlying
mass in stars has been recently explored.  \citet{kauffmann2003a} found
that the concentration of galaxies (a structural parameter correlated
with morphology) depends solely on the underlying stellar mass of the
galaxies, not on their environment.  \citet{kauffmann2004}
subsequently found that the star-formation rate depends inversely on
both the local galaxy density and the galaxy stellar mass.  These
results raise interesting questions about the relative dependence of
galaxy morphologies on galaxy mass and on local galaxy environment.

At higher redshifts, the star formation rate increases rapidly
\citep{lilly1996,madau98}.  Clusters of galaxies grow by the accretion
of galaxies and groups of galaxies, many of which will have a larger
star-formation rate than galaxies today.  Thus, we expect the fraction
of star-forming (late-type) galaxies to increase.
\citet{dressler97}, \citet{treu2003}, \citet{smith2005},
\citet{postman2005}, and \citet{desai2007} all found an increase in
the number of late-type galaxies at higher redshifts in clusters and a
decreasing fraction of S0 galaxies at redshifts from the present day
out to redshifts  $z\sim 0.5-1$.  This points to a substantial
buildup of cluster galaxies in the last half of the age of the
universe, i.e., over look-back times from 4 to 8 Gyr.  

Endeavoring to understand the details and the causes of these changes
is challenging, however, since luminosity was used to select the
galaxies.  Variations in the star formation rate can cause large
changes in the mass-to-light ratios of galaxies (these changes can
exceed an order of magnitude).  Galaxies of similar mass thus can be
spread across a range of luminosities (or be excluded altogether if
too faint).  Selection directly by the stellar mass of galaxies would
remove some of these problems. For example, mass-selected samples
would not include very low-mass galaxies undergoing
bursts of star formation.  These galaxies would enter
luminosity-selected samples, and distort comparisons between galaxy
populations at different epochs.  The stellar mass of galaxies changes
more modestly, even after a major mergers.

This has particular value for determining the progenitors of the
galaxy population in present-day clusters.  Galaxies of similar masses
at higher redshifts are likely to form the majority of the progenitors
of current galaxies.  Except for the rare major merger, mass growth
for the typical cluster galaxy is likely to be modest.  Merging and
star-formation will move galaxies to higher masses over time, so our
higher-redshift mass-selected samples will not contain all progenitors
of $z\simeq 0$ cluster galaxies.  However, these processes will
impact luminosity-selected samples much more dramatically.  Proper
accounting and understanding of the evolution of galaxy populations
(and of the nature of relations such as the morphology-density
relation) will be better done using mass-selected samples.

Given this, we set out to use mass-selected samples to investigate how
the morphological mix in clusters of galaxies changes with local
galaxy density and how the nix varies with time.  The aim of our 
investigation is to determine whether the trends found in the
luminosity--selected samples of \citet{dressler97}, \citet{smith2005},
and \citet{postman2005} appear in mass--selected samples.  The ultimate
goal is to establish the progenitors of the $z\simeq 0$ elliptical and
S0 population by examining the masses of $z\simeq 0 - 1$
galaxies.

We have constructed a sample of five clusters of galaxies; Coma at
$z=0.023$, CL~1358.4+6245 (hereafter \clt) at $z=0.33$, MS~2053.7$-$0449
(hereafter \mst) at $z=0.59$, MS~1054.4$-$0321 (hereafter \ms) and
RX~J0152.7$-$1357 (hereafter \cl), both at $z=0.83$.   Each cluster
has a large catalog of spectroscopic members, combined with high
resolution imaging and precise colors in the optical rest-frame.
These clusters are massive systems with both large X-ray luminosities
and substantial velocity dispersions.  This provides two advantages:
first, these rich systems provide large samples for studying the
distribution of morphologies; second, our cluster selection minimizes
the influence of the relation between X-ray luminosity or velocity
dispersion and early-type galaxy fraction found in \citet{postman2005}
and \citet{desai2007}.  For each cluster, we have rest-frame $B$
imaging, generally from the {\em Hubble Space Telescope} ({\em HST}),
which provides visual morphological classifications consistent with
those used in previous work such as \citet{dressler1980a,
dressler1980b, postman1984, treu2003, dressler97, smith2005,
postman2005, desai2007}.  Many of the classifications come directly
from the work of \citet{dressler1980b} and \citet{postman2005}.  The
underlying stellar mass is derived using the mass-to-light ratio
($M/L$) from the redshifted $B-V$ color \citep[see][]{kelson2000c,
bell2001, bell2003, holden2006} and total magnitudes from our
measurements.

We discuss our data in \S 2, with emphasis on how we select galaxies
to ensure that we have consistent morphologies and photometry across a
variety of clusters.  We discuss our mass estimates as well, with a
longer discussion in an Appendix \ref{masses}.  In \S 3, we turn to
our estimates of local galaxy density, showing that we reproduce the
observed evolution in the morphology--density relation reported by
other authors.  We then create mass--limited samples of galaxies in
each cluster and examine the evolution of the morphology--density
relation for galaxies that, at low redshift, are the massive
early-types that dominate cluster cores.  We discuss our results in
\S 4 with a summary in \S 5.

Throughout this paper, we assume $\Omega_m = 0.3$, $\Omega_{\Lambda} =
0.7$ and $H_o = 70\ {\rm km\ s^{-1}\ Mpc^{-1}}$.  All stellar mass
estimates are done using a ``diet'' Salpeter initial mass function
\citep[IMF][]{bell2001, bell2003}.  This IMF has the slope of a
traditional Salpeter IMF \citep[$x=1.35$]{salpeter1955} over the
range of 125$-$0.35 $M_{\sun}$, with a flat slope below
0.35 $M_{\sun}$.  This IMF produces the same colors and luminosity as
a Salpeter IMF, with 70\%, or 0.15 dex, of the $M/L$ of the original
Salpeter IMF \citep{salpeter1955}.  \citep{bell2001} selected this IMF
as it matches the properties of low redshift galaxies.  This IMF
yields masses 0.1$-$0.15 dex larger than the IMF of
\citet{kennicutt1983}, \citet{kroupa1993}, or \citet{chabrier2003}.

\section{Data and Object Selection}
\label{data}

We studied five clusters of galaxies in this paper, and they are
listed in Table \ref{summary}.  Each cluster in our sample is massive,
with a velocity dispersion $>850\ {\rm km\ s^{-1}}$, and is X-ray
luminous.  All but Coma and \cl\ were included as part of the {\em
  Einstein} Medium Sensitivity Survey cluster sample
\citep{gioia90a,henry92,gioia94}.  We list velocity dispersions and
bolometric X-ray luminosities in Table \ref{other}.

For each of the clusters, we used morphological types, photometry and
redshifts from the literature to compute the fraction of early-type
galaxies as a function of luminosity, mass and local galaxy density.
In columns (5)$-$(7) of Table \ref{summary}, we list the number
of galaxies used to estimate the local galaxy densities, the number
used to compute the MDR for luminosity-selected samples, and the
number of galaxies included in the mass-selected samples.  When
estimating the local galaxy density, we selected galaxies down to the
limiting magnitude of $M^{\star}_{V}+ 1.5$, where $M^{\star}_{V}$ is
the absolute magnitude from the \citet{schechter76} luminosity
function in the $V$ band.  We used the same value for $M^{\star}_{V}$
as \citet{postman2005}, $M^{\star}_{V}=-21.28$, and we used the same
evolution with redshift where $M^{\star}_{V} = M^{\star}_{V}(z=0) -
0.8 z$.  We chose a different luminosity-limit, $M^{\star}_{V}+ 1$,
for our MDRs This limit was simply selected to have about the same
size as a mass selection.  We selected a mass limit of $10^{10.6}
M_{\sun}$ as that is our completeness limit at $z=0.83$.  We 
explain this further in \S \ref{weights}.  Finally, we selected
galaxies within 1.25 Mpc of the cluster center, a limit imposed by the
{\em HST} imaging, except for the Coma Cluster where we extended the
data out to 1.5 Mpc for comparing our results with previous work in
the literature.

We would like to emphasize that in this paper, early-type galaxies are
morphologically--classified visually as elliptical and S0 galaxies
from the imaging data.  The classification was done using the image
with the filter closest to rest-frame $B$.  Neither the galaxy color
nor the spectrum was considered, thereby ensuring consistency with the
extensive body of previous work.  We made this distinction because
strongly bulge-dominated systems, such as elliptical and S0 galaxies,
have been shown to evolve differently from spiral galaxies.
Furthermore, galaxies selected based on spectral energy distribution
or color often show a mix of morphological types and star formation
rates \citep[see][for a recent example]{popesso2006}.

For all the galaxies in our clusters, we collected or determined
rest-frame optical morphologies, $B-V$ colors in the redshifted
passbands (hereafter $B_z - V_z$), total magnitudes and stellar mass
estimates.  We discuss below the data we used to construct these
samples.  


\begin{deluxetable*}{llrrccc}
\tablecolumns{7}
\tablecaption{Summary of Cluster Data}
\tablehead{\colhead{} & \colhead{} & \colhead{Observed} &
  \colhead{Redshift} & \colhead{No. of }  & \colhead{No. of} &
  \colhead{No. of} \\
\colhead{Cluster} & \colhead{$z$} & \colhead{ Filters} & \colhead{
  Range\tablenotemark{a}} & \colhead{All\tablenotemark{b}} &
\colhead{Luminosity\tablenotemark{c}} & \colhead{Mass\tablenotemark{d}} \\
}
\startdata
Coma & 0.023 & $u\ g\ r$ & $0.013 < z < 0.033$ &202 & 100 & 95 \\
\clt & 0.328 & $V_{606}\ I_{814}$ & $0.315 < z <
0.342$\tablenotemark{e} & 134 & 95 & 99 \\
\mst & 0.587 & $V_{606}\ I_{814}$ & $0.57 < z < 0.60$\tablenotemark{f}
& 85 & 60 & 56 \\
\ms & 0.831 & $V_{606}\ i_{775}\ z_{850}$ & $0.80 \le z \le
0.86$\tablenotemark{g} & 112 & 90 & 82\\
\cl & 0.834 & $r_{625}\ i_{775}\ z_{850}$ & $0.81 \le z \le
0.87$\tablenotemark{g} & 142 & 104 & 109 \\
\enddata
\tablenotetext{a}{The redshift range used to define cluster
  membership.  }
\tablenotetext{b}{The number of galaxies with redshifts in the range
  of the cluster with $M_V < M^{\star}_{V} + 1.5$.}
\tablenotetext{c}{The number of galaxies with redshifts in the range
  of the cluster with luminosities $M_V < M^{\star}_{V} + 1$.}
\tablenotetext{d}{The number of galaxies with redshifts in the range
  of the cluster with masses $> 10^{10.6} M_{\sun}$.}
\tablenotetext{e}{The catalog of \citet{fisher1998} contains only galaxies
  in this redshift range.}
\tablenotetext{f}{From \citet{tran2003}.}
\tablenotetext{g}{From \citet{blakeslee2005}.}

\label{summary}
\end{deluxetable*}

\begin{deluxetable*}{llll}
\tablecolumns{4}
\tablecaption{Summary of Ancillary Data}
\tablehead{\colhead{} & \colhead{} & \colhead{$\sigma$} &
  \colhead{$L_{X}$\tablenotemark{a}} \\
\colhead{Cluster} & \colhead{$z$} & \colhead{km s$^{-1}$} 
& \colhead{$10^{44}$ erg s$^{-1}$} \\
}
\startdata
Coma & 0.0231 & 1008 $\pm$ 33\tablenotemark{a} & 9.0 $\pm$ 0.2\tablenotemark{b}\\
\clt & 0.328 & 1027$^{+51}_{-45}$\tablenotemark{c} & 10.2 $\pm$ 0.7\tablenotemark{d} \\
\mst & 0.587 & 865 $\pm$ 71\tablenotemark{e} & 6.5 $\pm$ 0.4\tablenotemark{d}\\
\ms & 0.831 & 1156 $\pm$ 82\tablenotemark{f} & 16.4 $\pm$ 0.8\tablenotemark{g} \\
\cl & 0.834 & 919 $\pm$ 168\tablenotemark{h} & 18.6 $\pm$ 1.9\tablenotemark{i} \\
\enddata
\tablenotetext{a}{\citet{struble99}}
\tablenotetext{b}{\citet{voevodkin2004} }
\tablenotetext{c}{\citet{fisher1998}}
\tablenotetext{d}{\citet{henry2004}}
\tablenotetext{e}{\citet{tran2003}}
\tablenotetext{f}{\citet{tran2007}}
\tablenotetext{g}{\citet{gioia2004}}
\tablenotetext{h}{\citet{demarco2005a}}
\tablenotetext{i}{\citet{romer2000}}

\label{other}
\end{deluxetable*}

\subsection{Morphologies}

The morphological classifications came from the literature for all the
galaxies in our sample.  For the Coma Cluster, we used the
morphologies from \citet{dressler1980b} combined with imaging data
from the SDSS.  We selected early-type galaxies as those classified
as D, E, E/S0, S0/E, S0 and S0/a in \citet{dressler1980b}.  Late-type
galaxies are any galaxies with classifications in the spiral sequence
and irregular galaxies.  For all of the clusters in our sample, we used the
same classification criteria.

For \clt, we used the morphological types from \citet{fabricant2000}.
We selected early-type galaxies as those with a classification less than or
equal to 0.  This puts the Sa/S0 class in the early-type galaxy
category.  We used the unpublished data of \citet{tran2003,tran2004}
to form a sample of spectroscopic cluster members
with morphologies for \mst.  These classifications were done in the
same manner as in \citet{fabricant2000} and we used the same divisions
between early and late-type galaxies.  Finally, for \ms\ and \cl, we used the
previously published morphologies from \citet{postman2005} and
\citet{blakeslee2005}.  The classifications in \citet{postman2005} are
on a very uniform system, one consistent with the earlier samples of
\citet{fabricant2000} and \citet{dressler1980a}.

\subsection{Photometry}

We require a consistent set of magnitudes and colors to estimate
stellar masses.  We used the approach described in
\citet{blakeslee2005} to measure color and total magnitudes for the
cluster galaxies in our sample.  This procedure is a modification of
the approach outlined in \citet{vandokkum1998b}.  Each galaxy's
surface brightness distribution was fit by a S{\'e}rsic model with the
exponent ranging from $n=1$ to $n=4$, using the program {\tt galfit
}\citep{sersic,penggalfit2002}.  The fitted model is convolved with a
point-spread function, accounting for the different responses in our
data.  The fitted magnitude was used as our total magnitude estimate,
as we have found that to be a reliable measure
\citep{holden2004,holden2005b}.  We then applied the CLEAN
algorithm \citep{hogbom74} to the original images in all passbands.
We measured the flux within the half-light radius determined from the
surface brightness fit to measure the colors.  We detail below how the
colors were measured for each set of cluster galaxies since the
uniformity and consistency with which we carried out the photometry is
an important factor in ensuring reliable mass estimates across the
redshift range studied here.  Overall the agreement with other studies
was very good.

\subsubsection{Coma}

We created a mosaic of SDSS imaging covering all the galaxies in
\citet{dressler1980b}.  We fit the S{\'e}rsic model to the $g$
image. For a second estimate of the galaxy colors, we froze all of the
parameters of the model, except for normalization, and fit the surface
brightness profiles in the $u$ and $r$ data.  This is referred to as a
model color, and is used in the SDSS database.

We compared the total magnitudes with those from both \citet{jfk92}
and \citet{beijersbergen2002a,beijersbergen2002b}.  We found that our
total magnitudes were almost identical, a difference of $0.006 \pm
0.017$ from those of \citet{jfk92}. However, we found that the
\citet{beijersbergen2002a,beijersbergen2002b} magnitudes were 0.12 mag
fainter than our measurements.  The $r$ magnitudes from
\citet{beijersbergen2002a,beijersbergen2002b} are SExtractor
MAG\_AUTO magnitudes.  For the very brightest galaxies, we found that
the offset was larger, $\simeq 0.2-0.4$ mag, which we did not find
when we compared our total magnitudes to those of \citet{jfk92}.  When
we measured MAG\_AUTO magnitudes from our mosaic, we found a similar
offset for the brightest galaxies.  This probably results from the
large number of neighbors for the brightest galaxies in Coma,
suggesting that these measurements are not as reliable for total
magnitudes.

We compared our colors with those of the SDSS and those from
\citet{eisenhardt2007} and found very good agreement.  We found that our
colors, measured within a half-light radius, were redder by only
$0.013 \pm 0.004$ mag than the SDSS model colors.  We also compared
our model colors with those of the SDSS, and found that those were
redder by $0.006 \pm 0.005$ mag.  \citet{eisenhardt2007} give $B-V$
colors for a number of Coma members.  We computed a similar conversion
that in Table \ref{transform}, but for observed $B-V$ instead of
redshifted colors, $B_z - V_z$.  Using the relation $B-V = 1.00 (g-r)
+ 0.15$ to transform our half-light radius $g-r$ colors, we compared
our calculated $B-V$ colors with the measured $B-V$ colors in
\citet{eisenhardt2007}.  We found our colors are bluer by $-0.008 \pm
0.003$ with a scatter of 0.029 mag when compared with the colors
measured within $r_1$, a size that roughly corresponds to half of the
light, see \citet{eisenhardt2007} for details.

\subsubsection{\clt}

For every galaxy in the redshift catalog of \citet{fisher1998}, we
derived photometry from F606W ($V_{606}$) and F814W ($I_{814}$) Wide
Field Planetary Camera 2 (WFPC2)
imaging presented in \citet{vandokkum1998b}. We used the zero-points
from \citet{kelson2000a} for the WFPC2 photometry.  We used our
standard procedure, using the imaging in the $I_{814}$ passband to
determine the half-light radii.  The colors were measured in a similar
manner in \citet{vandokkum1998b}.  We found an offset of 0.018 $\pm$
0.006 mag from the \citet{vandokkum1998b} colors, and we found a
scatter of $\sigma(V_{606}-I_{814}) = 0.07$ mag between the two sets
of data.  Our total magnitudes are -0.021 $\pm$ 0.030 mag brighter
than the total magnitudes from \citet{kelson2000a}.

\subsubsection{\mst}

The photometry we used is based on the archival WFPC2 $V_{606}$ and
$I_{814}$ imaging.  \citet{wuyts2004} gives velocity dispersions and
the results of fitting surface brightness profiles to the WFPC2
imaging for a subset of the galaxies.  We measured the apparent
magnitudes for all cluster galaxies in the WFPC2 imaging using the
same procedure as for \clt. We compared our total magnitudes from surface
brightness profiles with those from \citet{wuyts2004}.  We find that
the $I_{814}$ magnitudes of \citet{wuyts2004} are fainter by 0.037
$\pm$ 0.020 mag than ours, so we consider the two sets of magnitudes
in good agreement.

\subsubsection{\ms\ and \cl}

\citet{blakeslee2005} measured the colors within the half-light
radius, with that radius determined from model fits, using the program
GALFIT \citep{penggalfit2002}. \citet{blakeslee2005} let the
S{\'e}rsic exponent range from $n=1$ to 4.  The total magnitudes
came from the SExtractor MAG\_AUTO parameter \citep{bertin96}.  To
determine total magnitudes, a 0.2 mag aperture correction was applied
to the measured magnitude \citep[see][]{holden2005b, blakeslee2005}.

\citet{blakeslee2005} only analyzed the then existing catalogs of
redshifts from \citet{demarco2005a} and \citet{tran2007}.  We have
acquired additional redshifts for both clusters since the publication
of \citet{blakeslee2005}. We used the Magellan telescopes with both
the Inamori-Magellan Areal Camera and Spectrograph
\citep{bigelow2003,dressler2006} and the Low-Dispersion Survey
Spectrograph 3.  These new members were analyzed in the same manner as
presented in \citet{blakeslee2005}.

For \ms, we also used the imaging data from the Faint Infrared
Extragalactc Survey
\citep[FIRES][]{forster2006}.  These data cover $UBVJHK$ along with the
archival WFPC2 photometry for \ms.  We replaced the WFPC2 data with
the Advanced Camera for Surveys (ACS) imaging and recreated the
catalogs.  In Appendix \ref{fires}, we discuss how we processed the
data.  For most filters, we found very small scatter, $<0.01$ mag,
between our measured magnitudes and those provided in the FIRES
catalogs.  Once we constructed a photometric catalog, we fit spectral
energy distributions to measure redshifted colors and the mass in
stars for cluster members.  We used these mass and color estimates as
a check for our masses and colors using only {\em HST} imaging.  We
discuss these in \S\S \ref{mag} and \S \ref{mass_est} below.

\begin{deluxetable*}{lrlr}
\tablecolumns{4}
\tablecaption{Photometric Conversions}
\tablehead{\colhead{Cluster} & \colhead{Redshifted Filter} &
  \colhead{Transformation}  & \colhead{Distance Modulus\tablenotemark{a}}\\ 
\colhead{} & \colhead{} & \colhead{} & \colhead{}\\
}
\startdata
Coma & $V_z$ & $ g - 0.65(g-r)+ 0.01$\tablenotemark{b} & 35.01 \\
Coma & $B_z - V_z$ & $ 1.00(g-r) + 0.15$\tablenotemark{b} & \\
\clt & $V_z$ & $ I_{814} + 0.27(V_{606}-I_{814}) + 0.63$ & 41.18 \\
\clt & $B_z-V_z$ & $ 0.91(V_{606}-I_{814}) - 0.19$ &  \\
\mst & $V_z$ & $ I_{814} - 0.21(V_{606}-I_{814}) + 1.07$ & 42.68 \\
\mst & $B_z - V_z$ & $ 0.70(V_{606}- I_{814}) - 0.30$ & \\
\ms & $V_z$ & $ z_{850} -0.16(V_{606}-z_{850})+0.74$\tablenotemark{b,c} & 43.60
 \\
\ms & $B_z- V_z$ & $ 1.05(i_{775}-z_{850})+0.11$\tablenotemark{b,c} &  \\
\cl & $V_z$ & $ z_{850} - 0.20(r_{625}-z_{850}) +
0.78$\tablenotemark{b} & 43.61 \\
\cl & $B_z - V_z$ & $ 1.05(i_{775}-z_{850}) + 0.11$\tablenotemark{b} &  \\
\enddata
\tablenotetext{a}{The distance modulus assumes $\Omega_m = 0.3$,
$\Omega_{\Lambda} = 0.7$, $H_o = 70\ {\rm km\ s^{-1}\ Mpc^{-1}}$. }
\tablenotetext{b}{ The Coma $gr$ data and the ACS photometry use an 
  AB zero point.}  
\tablenotetext{c}{The $V_{606}$ filter used in this
  conversion refers to the ACS F606W and not the WFPC2 filter used in
  \clt\ and \mst.}

\label{transform}
\end{deluxetable*}

\subsection{Redshifted Magnitudes and Colors}
\label{mag}

To compute the rest-frame magnitudes in $B_z$ and $V_z$ from the
observed magnitudes and colors, we used the technique of
\citet{blakeslee2005} and \citet{holden2006}.  We calculated the
magnitudes of templates in the rest-frame filters.  We then redshifted
the templates, and computed the magnitudes in the observed filters.
For the templates, we used exponentially decaying star formation rate
models from \citet[hereafter BC03]{bc03}.  These
models had exponential time-scales of 0.1 to 5 Gyr, covering a
range of ages from 0.5 to 12 Gyr and three metal abundances, 2.5
1 and 0.4 $Z_{\sun}$.  We list the
transformations between the observed filters and the redshifted ones
in Table \ref{transform} and include the distance modulus to the
cluster in our chosen cosmology.  For the rest-frame filters, we used
the $B$ and $V$ curves from \citet{buser1978} specifically the B3
curve for the $B_z$.  Both sets of curves are tabulated by BC03.

For \ms\ and \cl, $B_z$ lies within the observed wavelength range of
the ACS filters, while the center of $V_z$ is redshifted to reddest part the
\z\ filter at $z=0.83$.  To compute the MDR
for luminosity-selected samples as was done in previous work, we need
to compute the $V_z$ magnitude. We compared our $B_z- V_z$ colors for
\ms\ with those estimated from the broader baseline of photometry from
the FIRES survey.  We fit spectral energy distributions to the FIRES
data (see Appendix \ref{masses}).  We use those fits to interpolate
the observed colors to predict $B_z$ and $V_z$ for each galaxy.  We
found that the $B_z$ and $V_z$ had an average offset between the FIRES
predictions and the results from Table \ref{transform} of $\delta
(B_z-V_z) = -0.006$ mag with a scatter of $\sigma (B_z-V_z) = 0.005$
mag. When we compare the two estimates of the $V_z$ magnitude, we find
$\delta V_z = -0.033$ mag with a scatter of $\sigma V_z = 0.027$ mag.
Reassuringly, we find that our redshift colors and magnitudes from the
ACS imaging alone match what we expect from a much broader wavelength
coverage.

We have computed the transformations between the observed F606W and
F814W WFPC2 photometry and the redshifted $B_z$ and $V_z$ magnitudes
for \mst\ and \clt.  These are different conversions from those
previously listed in the literature.  For \clt, we found good
agreement between our transformations and those previously published
by \citet{vandokkum1998b}.  We found offsets of $\simeq$0.1 mag in the
colors of red galaxies in \mst\ between our transformations, listed in
Table \ref{transform}, and the published ones of \citet{tran2003}.
However, our transformation between the observed colors and $V_z$
agree with that of \citet{wuyts2004} to within $<0.05$ mag.  For
consistency across our data sets, we used the conversions we derived
that are listed in Table \ref{transform}.

\subsection{Mass Estimation}
\label{mass_est}

Estimating stellar masses requires a total magnitude and an estimate
of the mass-to-light ratio ($M/L$).  We used the redshifted colors and
magnitudes discussed in the previous sections to compute the total
magnitude in the $B_z$ filter, and we used the linear
relation between $M/L_B$ and the $B_z-V_z$ color given in
\citet{bell2003}, $M/L_{B} = 1.737 (B-V) - 0.942$, to derive $M/L$.
We assumed the absolute magnitude of the Sun to be $M_{B} =
5.45$.\footnote{This value of $M_{B}$ varies slightly from the value in
  \citet{binneym1998}; see
  http://www.ucolick.org/$\sim$cnaw/sun.html}  The relations of
\citet{bell2003} use a ``diet'' Salpeter IMF \citep[see][for an
explanation]{bell2001}.  This IMF has the same colors and luminosity
as a Salpeter IMF, but only 70\% of the mass.  The offsets between the
derived stellar masses assuming this IMF and other popular IMF's are
given at the end of \S 1 and in \citet{bell2003}.  Generally, most
other IMFs predict masses 0.1-0.15 dex lower than our estimates.

The assumption of a single relation between rest-frame color and $M/L$
has two likely sources of error.  First, there is an intrinsic scatter
in $M/L$ at a fixed color, which \citet{bell2003} estimate as 0.1-0.2
dex.  Second, the \citet{bell2003} $M/L$ estimates are appropriate for
$z=0$ stellar populations.  At higher redshifts, star-forming galaxies
may have lower stellar masses, as they have formed fewer stars, yet
they may have similar colors.  \citet{kassin2006} finds some evidence
that star-forming, field galaxies at higher redshift may have lower
stellar masses than galaxies with the same rest-frame colors at
$z=0$. \citet{kassin2006} find a shift of $\simeq$0.1 dex in the
stellar mass of star-forming galaxies at a fixed dynamical mass.  This
shift is larger, $\simeq$0.2 dex, if \citet{kassin2006} uses the
\citet{bell2003} stellar mass estimates.  For passively evolving
galaxies, \citet{vanderwel2006} find that systematic shifts of
$\simeq$0.1 dex are possible.  To estimate the sizes of the errors in
our masses, we computed separate $M/L$ estimates for a subset of our
data.  We also compared the $M/L$ from the \citet{bell2003} relations
with dynamical mass estimates in all of the clusters in our sample,
and we find a robust set of mass estimates that are consistent with
redshift.  The results of these tests are given below.

\subsubsection{Comparison of $M/L$ estimates}

For \ms, we fit spectral energy distributions to the photometric data
from the FIRES catalog.  In Appendix \ref{masses}, we discuss our
results from fitting spectral energy distributions (SEDs), using
exponentially declining star formation rate, or $\tau$, models, to the
data of \ms\ and a complementary sample of SDSS Fourth Data Release (DR4)
galaxies \citep{dr4}.  For both sets of data, we found that the
average $M/L_B$ agrees with the linear relation of \citet{bell2003},
see Figure \ref{ml_bmv_both}.  We found a scatter of 33\%, 0.14 dex,
in the $M/L$ around the best-fitting relation for \ms.  Early-type
galaxies have a very similar scatter of 34\%, or 0.15 dex, while
late-type galaxies have 30\%.  In our lower redshift sample created
using the SDSS DR4, we find a smaller scatter of 28\%, or 0.12 dex.
These error estimates agree well with the errors expected from
\citet{bell2003}.

\subsubsection{Comparison with Dynamical Masses}

\begin{figure}[htbp]
\begin{center}
\includegraphics[width=3.4in]{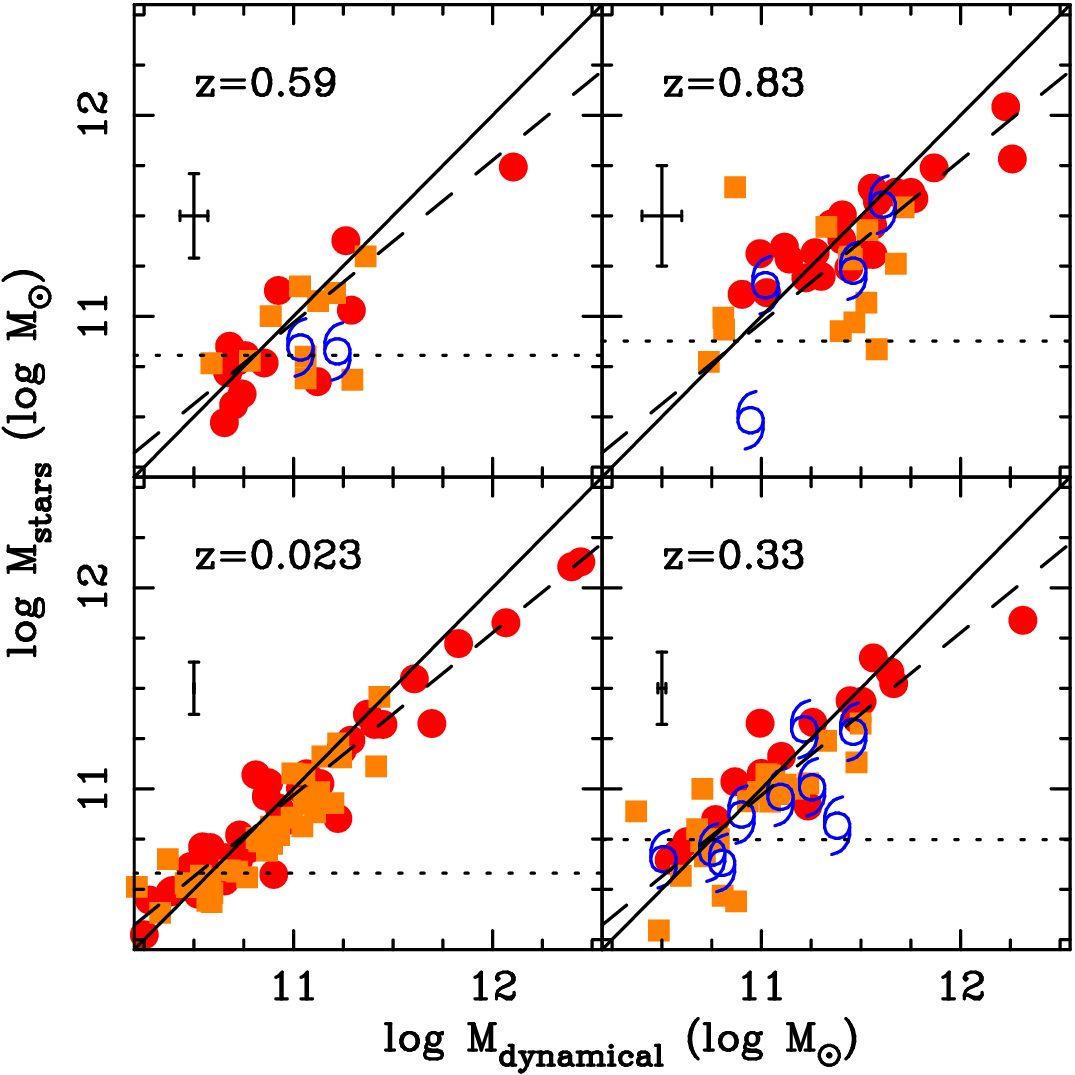}

\end{center}
\caption[f1.eps]{ Photometric stellar masses vs. dynamical masses.
  Red circles are elliptical galaxies, orange squares are S0 and S0/a
  galaxies, and blue spirals are later type galaxies.  Each cluster has a
  subset of galaxies with velocity dispersion measurements which we
  use to compute masses using the relation $M_{dyn} = 5
  r_{eff}\sigma^2 / G$. The dotted lines show the approximate mass
  limits for the velocity dispersion samples assuming the magnitude
  limits and the typical colors of the samples. The $z=0.83$ panel
  contains galaxies from two different clusters with two different
  magnitude limits; we show only the fainter limit. When computing
  statistics, we only use galaxies with masses above that line. The
  solid line has a slope of one and the dashed line shows the best
  fitting relation from \citet{gallazzi2006} using the de Vaucouleurs
  relation between size, velocity dispersion and mass.  We show the
  scatter around the relation of \citet{gallazzi2006} as a vertical
  error bar, and the average error for the dynamical masses as a horizontal
  error bar. It appears that, for the most massive galaxies, masses
  from the color estimates of $M/L$ may systematically, although
  slightly, underestimate the dynamical masses.  The average for each
  sample agrees with the \citet{gallazzi2006} intercept within
  $\pm$5\%.  Overall the photometric estimates provide a reliable and
  consistent estimate of the stellar mass in galaxies at all redshifts.}
\label{fpm}
\end{figure}

We compared our mass estimates with dynamical masses ($M_{dyn}$) from
fundamental plane measurements from the literature
\citep{jf94,jfk95sb,jfk95vel,kelson2000b,kelson2000c,wuyts2004,jorgensen2005}.
For each cluster, we combined the published velocity dispersions
($\sigma$) and sizes ($r_{eff}$) for the member galaxies using the
relation $M_{dyn} = 5 r_{eff} \sigma^2 / G$ \citep{jfk96}.  We find
that the relations between the dynamical mass, derived from the
velocity dispersion and $r_e$, and the stellar mass have consistent
zero-points, the difference was $\pm$0.02 dex, or $\pm$5\% for all of
the clusters in our sample (see Fig \ref{fpm}.)  The good agreement
between the dynamical and stellar masses at all redshifts shows that
we have a consistent mass scale for all early-type galaxies in our
sample, and one that matches that of others \citep{vanderwel2006,
rettura2006}.  

We do note that the slope between the two mass estimates appears to be
slightly less than one.  It seems as if the highest mass galaxies have
lower than expected stellar masses.  This is also seen in some other
comparisons \citep{gallazzi2006, rettura2006, kassin2006}.  We find
that we recover the same zeropoint for this relation as do
\citet{gallazzi2006}, after adjusting for the different IMF used in
that paper, over the same redshift range as \citet{rettura2006} or
\citet{kassin2006}.  We conclude that there is no systematic shift
with redshift in our stellar mass estimates, and that our photometric
stellar mass scale matches the dynamical mass scale very well.

When we compared the stellar mass estimates from our colors and total
magnitudes to the dynamical estimates, we found that the scatter
increased with redshift (see the vertical error bars in Figure
\ref{fpm}).  This increase occurs regardless of whether we compare with
the best-fitting relation, the relation of \citet{gallazzi2006} or a
simple slope of one.  For Coma, we found a scatter of 30\%, or 0.13
dex, close to the 0.17 dex scatter between color-based stellar masses
and dynamical ones quoted in \citet{vanderwel2006}.  At $z=0.83$, we
found a scatter of 58\%, or 0.25 dex, close to the scatter found at
$z\simeq 1$ by \citet{vanderwel2006}.  One possible reason for the
increased scatter is that the errors on the velocity dispersion
measurements increase.  If we remove the contribution from the
velocity dispersion error in the dynamical masses, the scatter only
reduces to 55\%, or 0.24 dex, so the increased scatter we find is not
from the statistical errors on the dynamical mass measurements.
\citet{kassin2006} find a scatter of 0.16 dex between the stellar
mass and the dynamical mass at all redshifts.  \citet{kassin2006} use
relations between color and $M/L$ that are different from ours, but they
recover the same slope and zeropoint as we do in Figure \ref{fpm}.
The difference in the scatter found by \citet{kassin2006} and our
results most likely comes from the details in how the measurements are
made.  \citet{kassin2006} explicitly models
a rotational component for later-type galaxies, which they found to
reduce the scatter.  Both \citet{gallazzi2006} and \citet{kassin2006}
find that the slope of the relation between the dynamical mass and the
stellar mass is less than one.  Using that best-fitting slope from
\citet{gallazzi2006}, we reduce the measured scatter to 25\%, or 0.11
dex, for Coma and 46\%, or 0.20 dex, for our two $z=0.83$ clusters.
Such scatter is well in line with the expectations from
\citet{bell2003}.

\subsection{Completeness Corrections and Catalog Limits}
\label{weights}

Each data set we used has a different completeness limit for the
redshifts and morphological identifications.  For each cluster in our
sample, we needed to give a weight to each galaxy based on the
probability that it would have either a redshift or a morphological
type, based on the galaxy's observed magnitude, hereafter $w(m,T)$.
The same approach was used by \citet{postman2005}.

For \ms, \cl, and \clt, the spectroscopic redshift catalogs were less
complete than our morphological catalogs.  Thus, we weighted the
galaxies with redshifts based on the fraction of objects with
redshifts as a function of morphological type and magnitude, as was
done in Appendix B of \citet{postman2005}.  The weight $w(m,T)$,
is inverse of the fraction of galaxies with redshifts, $N(z,m,T)/N(m,T)$
where $N(m,T)$ is the number of objects with classifications and
$N(z,m,T)$ is the number of objects with classifications and
redshifts.  Both fractions are calculated in a magnitude bin $ m_h \ge
m \ge m_l$ (we selected bins of 0.5 mag in size).

The data for \mst\ have the inverse problem, namely complete redshift
catalogs, but incomplete classifications.  We therefore weighted each
galaxy by the fraction of objects classified by the magnitude and
type, $w(m,T)$, such that \( w(m,T) = N(m_h \ge m \ge m_l, z, T
)/N(m_h \ge m \ge m_l, T) \) where $N(m,z,T)$ is the number of
late-type or early-type galaxies in a magnitude bin $m_h \ge m \ge m_l$ with
redshifts and $N(z,m)$ is the total number objects in that bin with
redshifts.  

\begin{figure}[htbp]
\begin{center}
\includegraphics[width=3.4in]{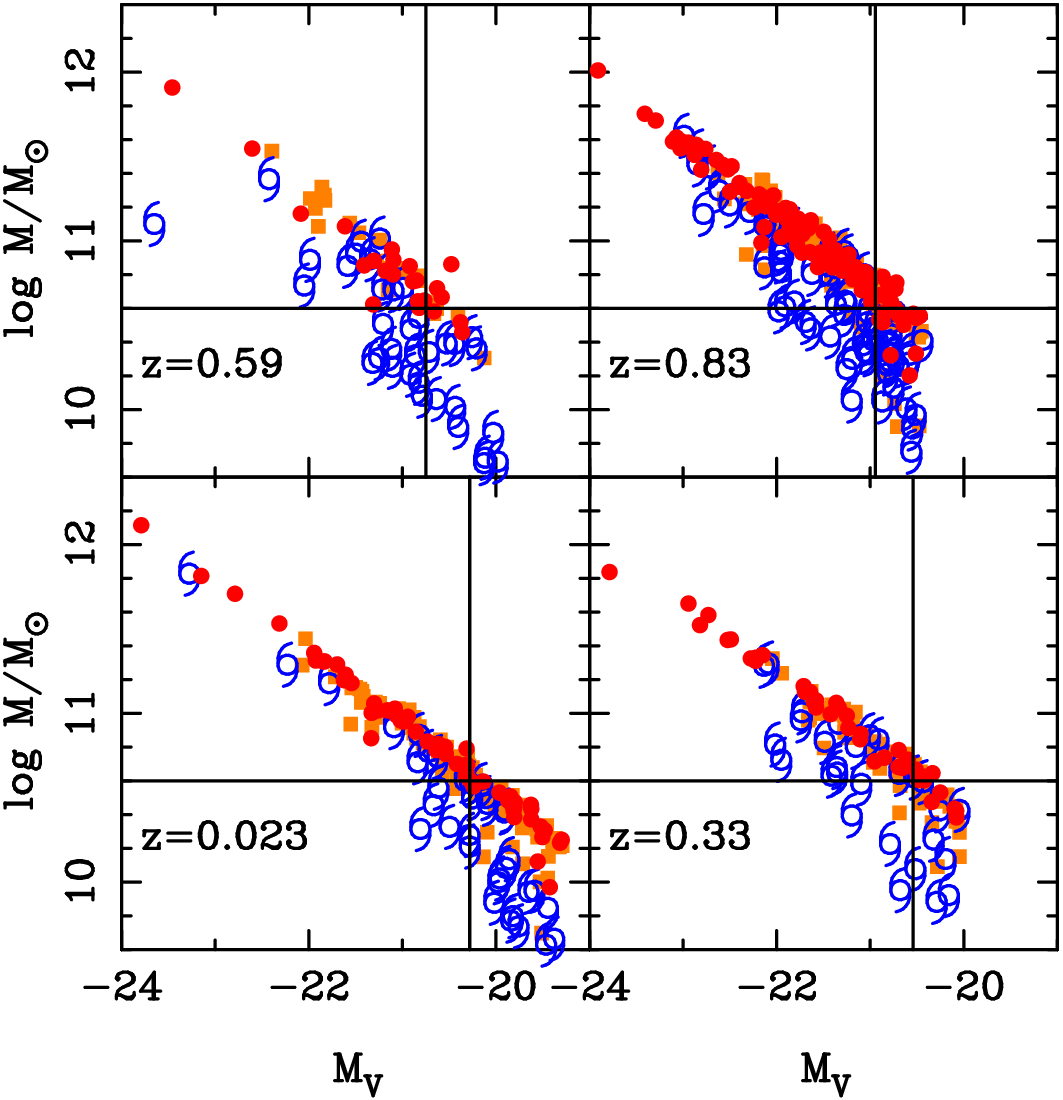}

\end{center}
\caption[f2.eps]{Photometric stellar masses as a function of absolute
  $V$ magnitude.  Red circles are elliptical galaxies, orange squares
  are S0 and S0/a galaxies, and blue spirals are later types.  The
  cluster samples, are Coma (bottom left), \clt\ (bottom right), \mst\
  (top left) and the combination of \ms\ and \cl\ (top right).
  The horizontal line is $M=10^{10.6}\ M_{\sun}$, the completeness
  limit for mass-selected samples.  They are derived by determining
  where the completeness if 50\% for the reddest galaxies at $z=0.83$
  (see \S \ref{weights} for a complete explanation).  The vertical
  line is the magnitude limit, $M^{\star}_{V} + 1$, we used for
  luminosity-selected samples, where $M^{\star}_{V} = -21.28 - 0.8 z$
  \citep{postman2005}.}
\label{mass_lum}
\end{figure}

We plot, in Figure \ref{mass_lum}, the stellar mass estimates as a
function of absolute $V$ magnitude for the five clusters in our
sample. To create a sample with a uniform-selection across all
redshifts, we selected a mass limit where we were 50\% complete at
$z=0.83$.  That limit is $M=10^{10.6} M_{\sun}$ for our choice in IMF
($M \simeq 10^{10.45} M_{\sun}$ for the \citet{kennicutt1983},
\citet{kroupa1993}, or \citet{chabrier2003} IMF's).  

\subsection{Projected Galaxy Density}

To estimate the projected galaxy density we used the same methodology
as \citet{postman2005}, i.e., a $n=7$ nearest neighbor distance.  We
included all galaxies within 1.5 Mpc of the cluster center with $M_V <
M^{\star}_{V} + 1.5$.  In total, there are \ntot\ galaxies in all five
clusters at or above this magnitude limit.

The local density was calculated to be \[ \Sigma = \frac{1}{\Omega_N
D_A^2} \sum^{7}_{n=1} {w_n(m_n,T_n)^{-1} } \] where $\Omega_N$ is the
angular area to the $N$th nearest neighbor, $D_A$ is the angular
diameter distance for the redshift of the cluster, and $w_n(m_n,T_n)$
are the weights for galaxy $n$ with a given magnitude $m$ and a
classification $T$.  We discuss how the weights are calculated in \S
\ref{weights}.

Because the weights are used in estimating both our densities and our
fractions by galaxy type, we derived errors for our density estimates
by bootstrapping.  We found that the errors on the density estimates
ranged from 0.17 dex for Coma to 0.24 dex for \ms .  Therefore, when
we computed the MDR, we used bin sizes of at least 0.5 dex.

\citet{dressler1980a}, \citet{dressler97}, and \citet{postman2005} all
used a fainter magnitude limit, $M^{\star}_{V} + 2$, than we did when
computing the densities.  We selected a brighter magnitude limit to
minimize the impact of sample incompleteness.  For comparison with
these previous results, we needed to correct our densities.
\citet{postman2005} had a similar problem. Since their data were not
always complete at that magnitude limit ($M^{\star}_{V} + 2$),
\citet{postman2005} computed a correction factor for their densities.
This correction factor was the ratio of the number of galaxies in a
Schechter luminosity function with $\alpha =-1.22$ down to
$M^{\star}_{V} + 2$ over the number of galaxies in the same luminosity
function down to the magnitude limit of the data.  Using a similar
procedure, and assuming the same luminosity function, the densities
for a magnitude limit of $M^{\star}_{V} + 2$ would be larger than our
densities by 0.18 dex.  In order to compare our results with
\citet{dressler1980a}, \citet{dressler97}, and \citet{postman2005}, we
have rescaled our densities by 0.18 dex.

\section{The Morphology-Density Relation and the Evolution in the
  Early-type Galaxy Fraction}

We combined the data discussed above and created two samples for each
cluster in order to derive MDRs and to study the evolution in the
early-type galaxy fraction with redshift. First, we replicated the
previous work using luminosity-selected samples, then we selected
galaxies using the masses estimated above.  For each sample, we
estimated the MDR in a consistent manner, using the same sets of
densities for both luminosity- and mass-selected samples.  In the
following subsections, we outline how we derived MDRs for our clusters.
   
\subsection{Early-type Fraction Estimates}

The early-type fractions for the clusters were not simply based on an
integer number of objects.  Instead, the early-type fraction was the
sum of weights of elliptical and S0 galaxies above our magnitude limit
or mass limit, divided by the sum of the weights of all galaxies
meeting that same limit.  In general, these weights are close to one,
and not using these weights changes the resulting fractions by only a
few percent. While the changes are small we felt it important to be
consistent and thorough in our approach.  The errors on the fractions
range from 3\% to 15\%, depending on the number of galaxies.  These
errors are close to, but slightly larger, than those expected from
using Poisson errors.  We added in quadrature an error of 6\% on the
early-type fractions we find for \cl\ and \ms.  This is the rms
classification error found by \citet{postman2005}.

\begin{figure}[htbp]
\begin{center}
\includegraphics[width=3.4in]{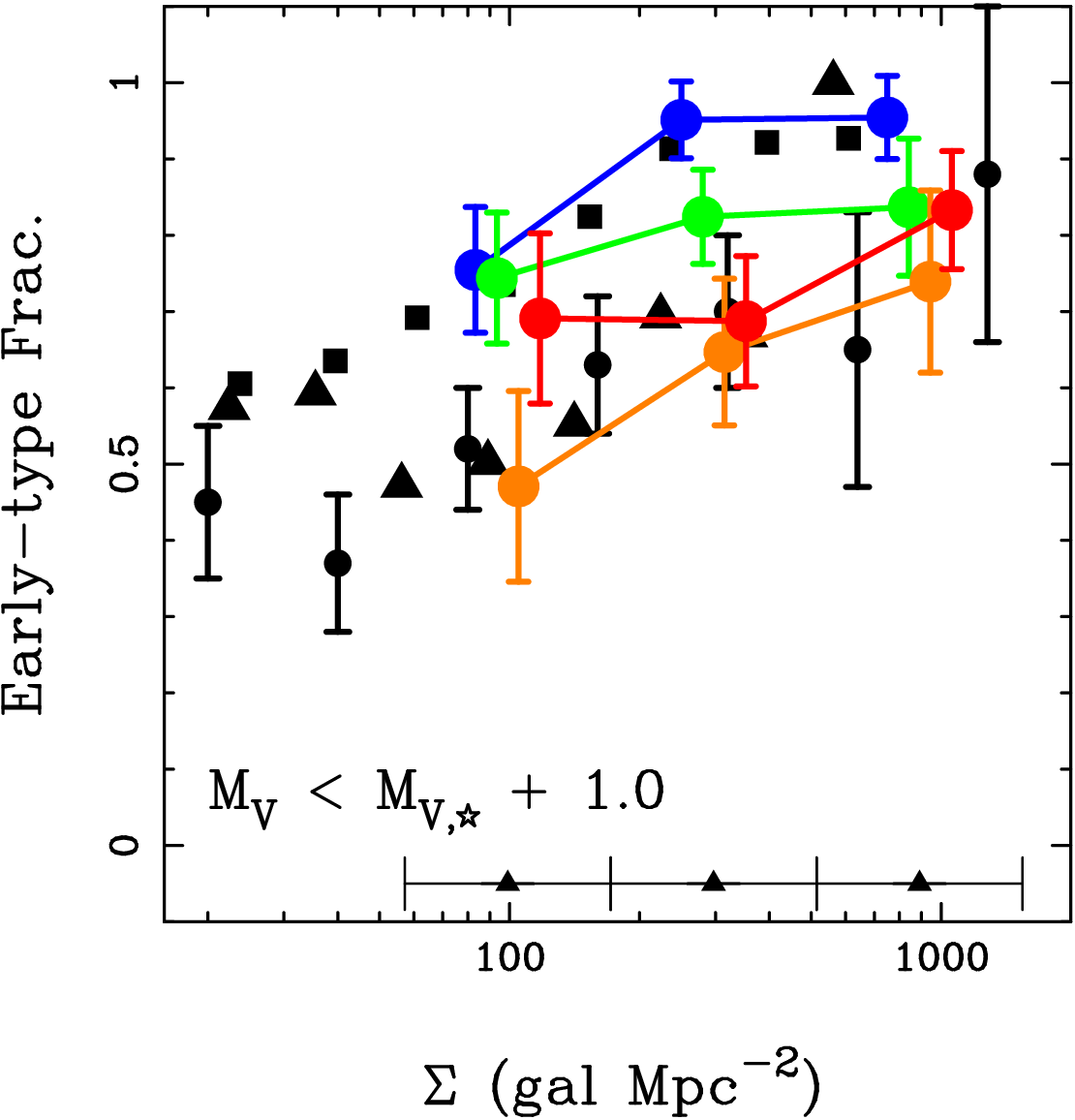}
\end{center}
\caption[f3.eps]{ Morphology-density relation for luminosity-selected
  galaxies by cluster. We selected all galaxies to an absolute $V$
  magnitude of $M^{\star}_{V} + 1$ or $M_V = -20.28-0.8 z$,
  \citep[see][]{postman2005}, for a total of \nlum\ galaxies in all
  five clusters. The cirlces connected by lines represent Coma (blue
  circles), \clt\ (green circles), \mst\ (orange circles) and the
  combined sample of \ms\ and \cl\ (red circles).  The bins are shown
  along the bottom, with triangles marking the centers. The galaxy
  fractions are derived in the same density bins, but are spread out
  for clarity around the bin centers in redshift order.  The data for
  Coma consist of the morphologies from \citet{dressler1980b} but
  photometry from the SDSS.  For comparison, the updated
  \citet{dressler1980a} results -- from \citet{dressler97} -- are
  shown as squares, the results of \citet{dressler97} at $z
  \simeq 0.4$ are shown as triangles, and the \citet{postman2005}
  results at $z\simeq 1$ are shown asblack circles with error bars.  We
  find good agreement between our morphology-density relations and the
  previous results from the literature. A trend in the early-type
  fraction with redshift is apparent, and is quantified further in
  Figure \ref{fraclz}.  We list the values for this figure in Table
  \ref{mdrs_lum}.}
\label{mdr_lum}
\end{figure}

\subsection{Luminosity-Selected Samples}
\label{lumsel}

\begin{figure}[htbp]
\begin{center}
\includegraphics[width=3.4in]{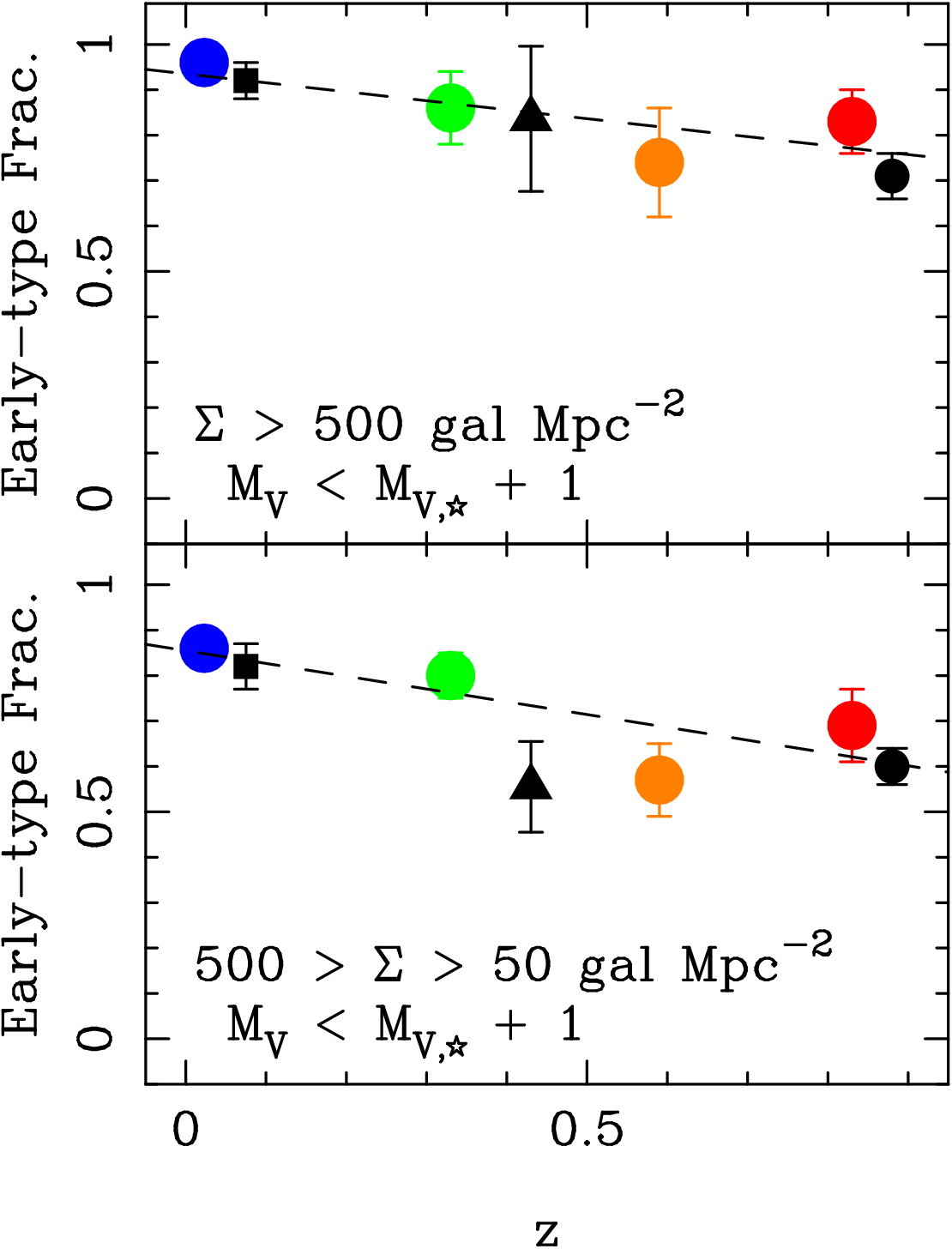}

\end{center}

\caption[f4.eps]{ Early-type fraction as a function of redshift in two
density bins for luminosity-selected samples. The density bins are
listed on the figure in galaxies Mpc$^{-2}$ (and differ from the bins used
in Figure \ref{mdr_lum}).   Blue, green, orange, and red circles
represent Coma, \clt, \mst\ and the two $z=0.83$ clusters,
respectively.  For comparison, the squares at low redshift are the
average fractions in the listed density bin from \citet{dressler97},
an updated version of \citet{dressler1980a}, the triangles show the
average fraction at $z\simeq 0.4$ from \citet{dressler97},  and the
black circles represent the high-redshift fractions from
\citet{postman2005}.  We note that these are not weighted by the
number of galaxies at each density, but are simply the average
early-type fraction inside the range of densities listed in the panel.
The dashed line represents a fit to the data from this paper, see
Table \ref{results}.  We see a strong trend in the evolution of the
early-type fraction with redshift, and a lower early-type fraction at
lower densities at most redshifts.  The evolution with redshift in the
early-type fraction is significant at the $>95$\% confidence limit
and is similar to that see by others. 

}

\label{fraclz}
\end{figure}

Our first step was to create luminosity-selected samples to compare
with previous work on the MDR. As discussed in \S \ref{data}, we
computed the morphology-density relation down to a magnitude limit of
$M^{\star}_{V}+ 1$ for all clusters, the same limit as was selected by
\citet{smith2005}.  We plot the resulting morphology-density relations
for a sample of a total of \nlum\ galaxies in Figure \ref{mdr_lum}.
We tabulate the plotted fractions in Table \ref{mdrs_lum}.  We see MDRs in
our clusters. Our data clearly shows a decrease in the early-type
fraction with decreasing density for luminosity-selected samples.
This evolution is consistent with that shown in previous work such as
\citet{dressler97}, \citet{smith2005} and \citet{postman2005}.
 
\subsubsection{Evolution of the Luminosity-Selected Early-type Fraction}

The MDRs in Figure \ref{mdr_lum} also show a clear trend of decreasing
early-type fraction with redshift.  To quantify this we examined the
fraction of early-type galaxies in two density bins as a function of
redshift, $\Sigma > 500$ galaxies Mpc$^{-2}$ representing high galaxy
densities, and $500 {\rm galaxies Mpc^{-2}} > \Sigma > 50$ galaxies
Mpc$^{-2}$ for low densities.  The cutoff of $\Sigma=500$ galaxies
Mpc$^{-2}$ was chosen to split the combined \ms\ and \cl\ into two
equal-sized subsamples.  The lower limit is the density that contains
95\% of our sample.  We plot our early-type fractions in each of these
density bins as a function of cluster redshift in Figure \ref{fraclz}.
We find a trend of a decreasing early-type fraction with redshift, as
seen before \citep{dressler97, smith2005, postman2005}.  For
comparison, we show the average fractions in the same density bins
from the low-redshift sample of \citet{dressler1980a}, updated in
\citet{dressler97}, the \citet{dressler97} $z\simeq 0.4$ sample and
the $z\simeq 1$ sample of \citet{postman2005}.  Our data agree with
the trend shown in the other published work, given the size of our
sample.

The dashed line in Figure \ref{fraclz} is a fit to the data that
includes errors on the estimated early-type fraction.  We find slopes
that differ from zero by more than 2$\sigma$ for both
fits; i.e., the non-zero slope is significant at the 95\% confidence
limit.  At high densities, we find that the early-type galaxy fraction
decreases by -0.20$z\ \pm$ 0.10 $z$, evolution at the 95\% confidence
limit.  For the low-density sample we find -0.29$z\ \pm$ 0.10 $z$,
evidence at the 99\% confidence limit for evolution.  Generally,
studies find that most of the evolution at low to moderate redshifts
happens at lower densities \citep{treu2003,smith2005,desai2007}, a
result we replicate with our data.  Our results at all densities are
clearly consistent with the trend seen from the previous work of
\citet{dressler1980a}, \citet{dressler97} and \citet{postman2005}.
We list the fractions in the columns (3) and (4) of Table \ref{results}.

\subsection{Mass-Selected Samples}
\label{mass_sel}

\begin{figure}[htbp]
\begin{center}
\includegraphics[width=3.4in]{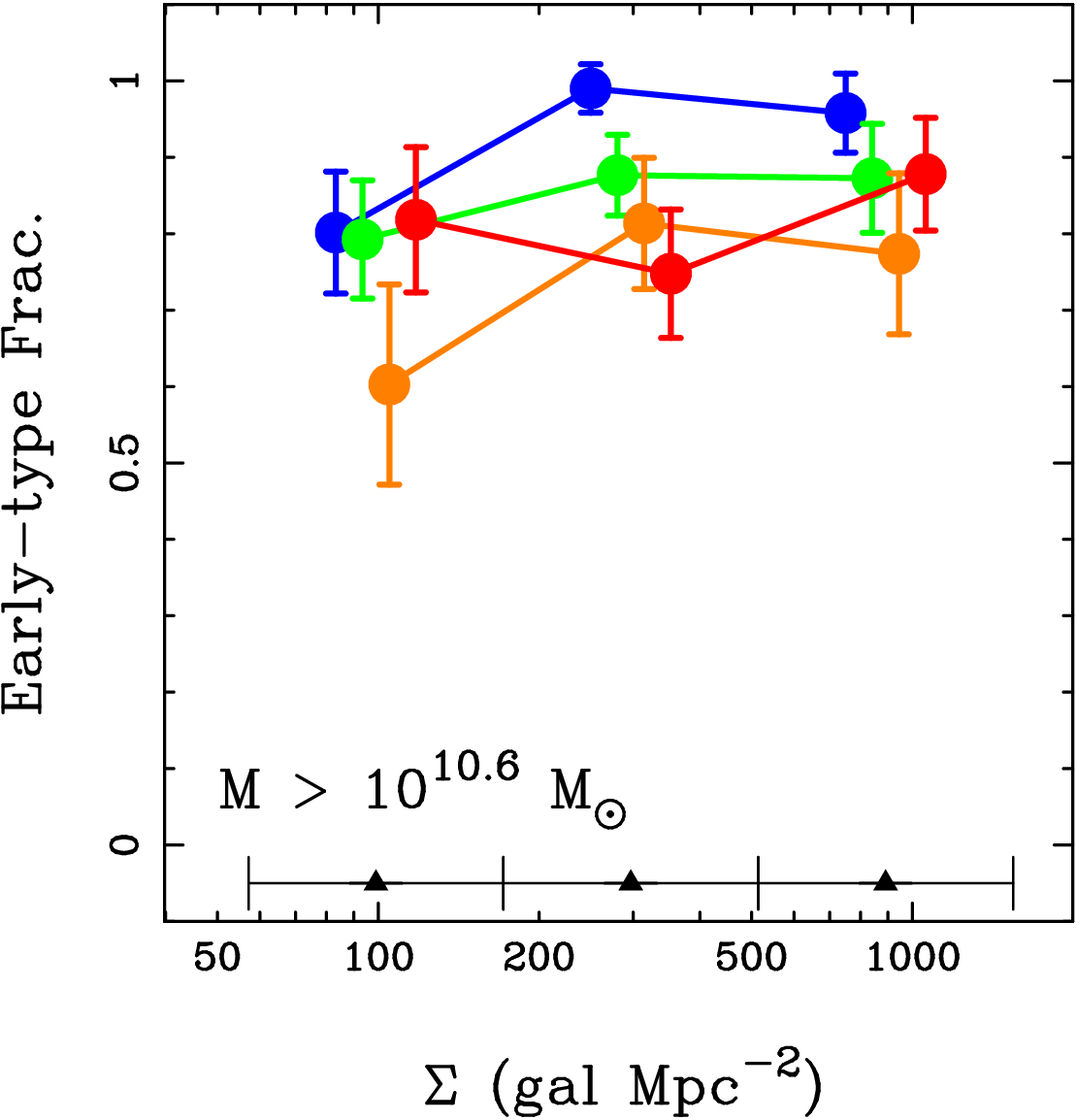}
\end{center}

\caption[f5.eps]{ Morphology-density relations for mass-selected
samples of galaxies in our clusters. The sample consists of all
galaxies in our five clusters above a mass limit of $>10^{10.6}
M_{\sun}$ \citep[$>10^{10.45} M_{\sun}$ for the IMF's
from][]{kennicutt1983, kroupa1993, chabrier2003}, or 0.25 ${\cal
M^{\star}}$ for a total of \nmass\ galaxies in all five clusters.  Our
mass-selected sample thus extends significantly below ${\cal
M^{\star}}$ (and ${ L^{\star}}$, since the mass limit corresponds
to 1 mag below ${ L^{\star}}$ for the reddest cluster galaxies),
and includes $\sim 70$\% of the stellar mass in galaxies in the
cluster.  Th symbols are the same as in Fig. \ref{mdr_lum}.  The
samples of galaxies at $\Sigma < 50$ galaxies Mpc$^{-2}$ are too small
to be statistically significant.  In contrast with Fig. \ref{mdr_lum}
the early-type fraction is larger, and the morphology-density relation
is less clear.  The trend with redshift also is noticeably less; see
Figure \ref{fracz}.  We list the values for this figure in Table
\ref{mdrs_mass}. 
}

\label{mdr_mass}
\end{figure}

\begin{figure}[htbp]
\begin{center}
\includegraphics[width=3.4in]{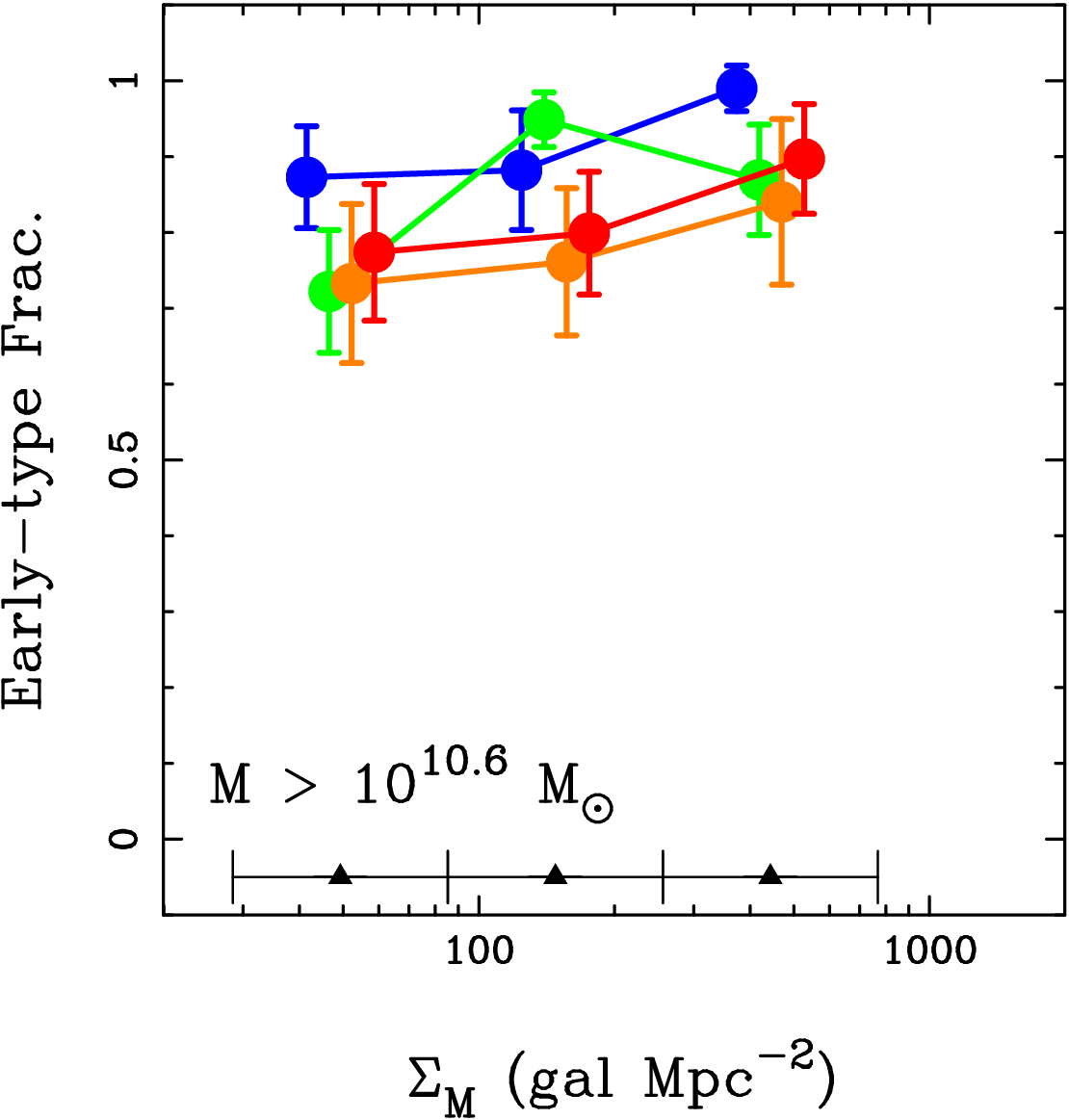}
\end{center}

\caption[f6.eps]{ Morphology-density relation for mass-selected
samples of galaxies using mass-selected galaxy densities. The symbols
are hte same as in Fig. \ref{mdr_lum}.
This figure is similar to Figure \ref{mdr_mass}, where we plot the
fraction of elliptical and S0 galaxies above the mass limit of
$>10^{10.6} M_{\sun}$.  In this figure, the local galaxy densities are
computed from the mass-selected sample ($\Sigma_M$), not the
luminosity-selected sample as in Figure \ref{mdr_mass}.  The change in
our estimate of the local galaxy density does not change our results.
The typical values for $\Sigma_M$ are a factor of $\simeq$2 smaller
than from our $M_V < M^{\star}_{V} + 1.5$ luminosity-limited sample,
which \citet{vanderwel2007a} finds as well for field galaxies.

}
\label{mdr_massd}
\end{figure}

\begin{figure}[htbp]
\begin{center}
\includegraphics[width=3.4in]{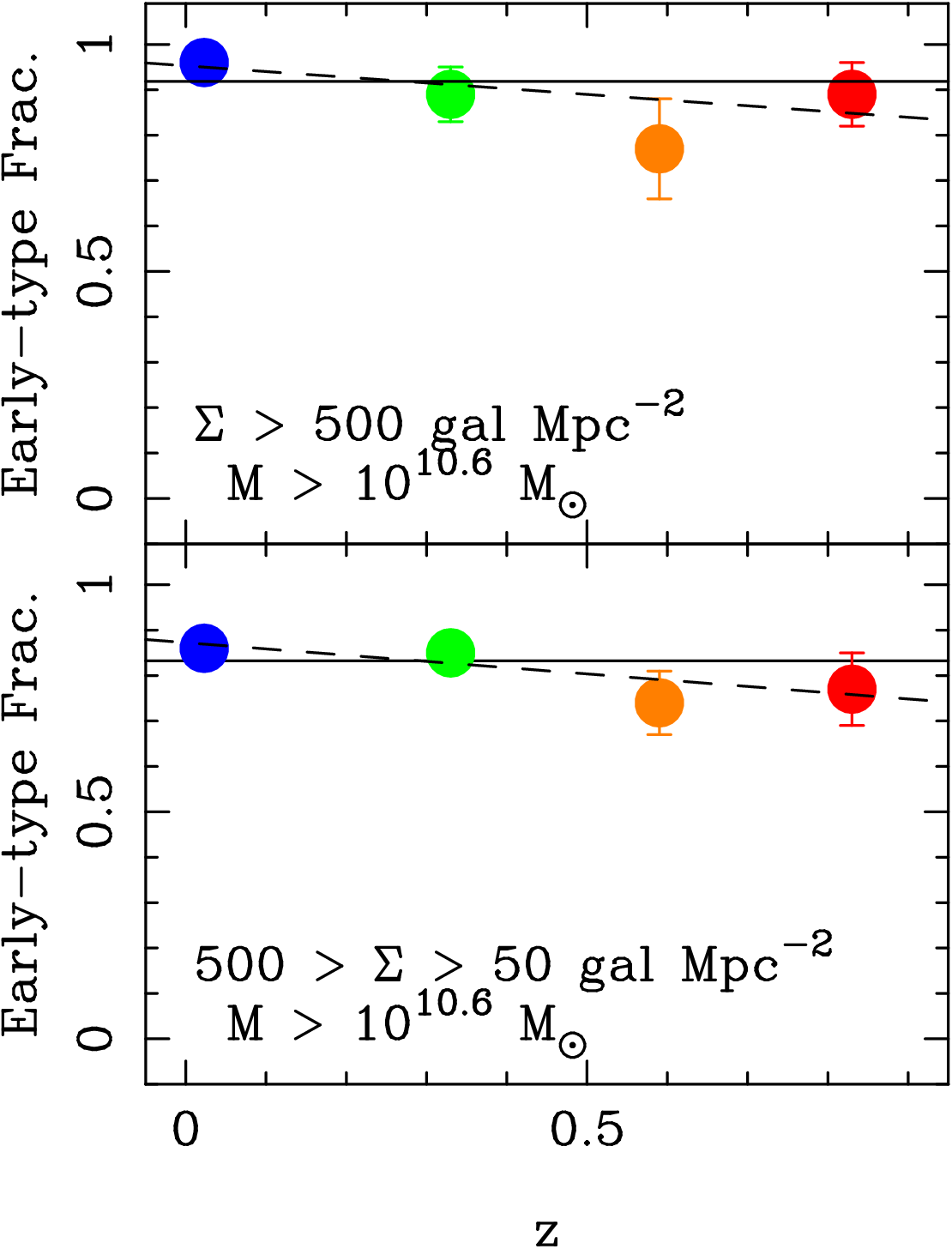}
\end{center}

\caption[f7.eps]{ Early-type fraction as a function of redshift for
  mass-selected samples in two density bins.  The symbols are the same
  as Fig.\ \ref{fraclz}.  The density range in ${\rm galaxies\ Mpc^{-2}}$
  and the mass limit are labeled in each plot.  We show the average
  early-type fraction in each bin as a solid line, with the height of
  the bar showing the error on the average.  We find mean early-type
  fractions of \ldf\ $\pm$ 3\% at $ 500\ > \Sigma > 50\ {\rm galaxies\
    Mpc^{-2}}$ and \hdf\ $\pm$ 4\% at $\Sigma\ > 500 \ {\rm galaxies\
    Mpc^{-2}}$. The dashed line is the best-fitting linear relation to
  the change in fraction with redshift, as in Fig.  \ref{fraclz}.  We
  list the fractions in Table \ref{results}. There is little evidence
  for evolution in the morphology-density relation above the mass
  limit, with the best-fitting slopes half or less than half those of
  Figure \ref{fraclz}.  The evolution in the luminosity-selected
  samples must occur predominately in galaxies with $M < 10^{10.6}
  M_{\sun}$.

}

\label{fracz}
\end{figure}

\subsubsection{The Mass-Selected MDR}

The next step was to determine the MDR for mass-selected samples. We
used the same weights and densities as in Figure \ref{mdr_lum} and \S
\ref{lumsel}.  Our redshifts or morphological classifications became
incomplete at $M < 10^{10.6}\ M_{\sun}$, so we select this as our mass
limit. There are a total of \nmass\ galaxies above that mass limit in
all five clusters. This mass limit corresponds to a factor of 4
below  $z=0$ ${\cal M^{\star}}$ for high-density regions
\citep{baldry2006}\footnote{As noted above, we use an IMF that adds
  0.15 dex to the masses of galaxies, raising the value of $\log {\cal
    M^{\star}}$ from 11.05 to 11.2. }, assuming no evolution in the
mass function.  If we integrate over this mass function, our mass limit
encompasses 68\%-73\% of the total stellar mass of the cluster
population, depending on the mass limit for the galaxy population
($10^{8}$ or $10^{9.5} M_{\sun}$, respectively).  To compute
this, we used the best-fitting mass function from \citet{baldry2006}
for high-density regions, i.e., $\Sigma>10$ galaxies Mpc$^{-2}$.  Thus
our mass-limited sample encompasses a wide range of masses for cluster
galaxies, and includes the majority of the stellar mass in galaxies in
clusters.

We show the resulting mass-selected MDR in Figure \ref{mdr_mass}.  The
mass-selected MDR (Figure \ref{mdr_mass}) is different from the
luminosity-selected MDR (Figure \ref{mdr_lum}).  First, the slopes
appear to be slightly less.  Second, at low redshift the MDRs appear
roughly comparable, but at higher redshift, at a given density, the
mass-selected early-type fraction is 10\%-20\% higher than the
luminosity-selected fraction. The contrast is quite striking,
considering that our mass sample extends significantly below ${\cal
  M^{\star}}$ (and below ${ L^{\star}}$, since the mass limit
corresponds to $\simeq 1$ mag below ${ L^{\star}}$ for the reddest
cluster galaxies).  See Tables \ref{mdrs_lum} and \ref{mdrs_mass} for
the resulting fractions in comparison with those from
luminosity-selected samples.

The MDR we show in Figure \ref{mdr_mass} is somewhat of a hybrid. We
estimated the galaxy densities using magnitude-limited samples, while
the early-type fractions came from mass-limited samples. We used a
luminosity-selected sample to estimate the local galaxy to minimize
the changes in comparing the two samples (luminosity-selected and
mass-selected).  In Figure \ref{mdr_massd}, we show the morphological
fractions as a function of a galaxy density estimated using a
mass-selected sample.  The mass limit we used was $10^{10.6} M_{\sun}$
and we used the same weights as discussed in \S \ref{weights}.  We
find that the mass-selected local galaxy densities are generally a
factor of $\simeq$2 lower than those from luminosity-selected
samples for the same galaxy, similar to that found by
\citet{vanderwel2007a} for field galaxy samples.  This offset in the
density does not, as expected, change the overall shallow slope, we
find the same value for the slope as we do in Figure \ref{mdr_mass}.
When plotting against these new densities, we still find a lack of
evolution as we found in Figure \ref{mdr_mass}.

The mass-selected MDR appears more shallow than the
luminosity-selected relation.  We tested this in a simple way.  We fit
the data from \citet{dressler1980a} with a linear slope in the
density range our sample covers.  We compared that slope to the
average slope for the MDR of all five clusters in our sample that we
plot in Figures \ref{mdr_lum} and \ref{mdr_mass}. The measured slopes
are $0.26 \pm 0.09$ for \citet{dressler1980a} and $0.25 \pm 0.13$ for
\citet{postman2005}.  Our mean luminosity-selected MDR has a slope of
$0.18 \pm 0.07$, while our mean mass-selected MDR have a mean slope of
$0.12 \pm 0.08$.  To measure these slopes, we average the samples
after removing the offset between the average early-type fractions for
each cluster.  This preserves the shape of the MDR for each sample.
The slope in the mass-selected MDR appears to be less than our
luminosity-selected MDR, but it is not statistically-significant.  We
find a larger, but still not statistically-robust difference between
the slope of the relation from \citet{dressler1980a} and
\citet{postman2005}, and our mass-selected sample.

Overall, the mass-selected MDRs do not differ substantially from the
luminosity-selected MDRs, except in one very important aspect -- their
evolution with redshift.

\begin{deluxetable*}{lllll}
\tablecolumns{5}
\tablecaption{Luminosity-Selected Morphology-Density Relations}
\tablehead{\colhead{Cluster} & \colhead{} &  
  \colhead{$1580> \Sigma > 500$\tablenotemark{a}} &
  \colhead{$500 > \Sigma > 167$\tablenotemark{a}} &
  \colhead{$167> \Sigma > 50$\tablenotemark{a}}  \\
\colhead{} & \colhead{} &  
  \colhead{${\rm galaxies\ Mpc^{-2}}$} &
  \colhead{${\rm galaxies\ Mpc^{-2}}$} &
  \colhead{${\rm galaxies\ Mpc^{-2}}$} \\
\colhead{} & \colhead{$z$} & \colhead{(\%)} &
  \colhead{(\%)} & 
  \colhead{(\%)} \\
}
\startdata 
Coma & 0.023 & 95 $\pm$ 5 & 95 $\pm$ 5 & 75 $\pm$ 8 \\
\clt & 0.328 & 84 $\pm$ 9 & 82 $\pm$ 6 & 74 $\pm$ 9 \\
\mst & 0.587  & 74 $\pm$ 12 & 65 $\pm$ 10 & 47 $\pm$ 12 \\
\ms\ \& \cl & 0.83 & 83 $\pm$ 8 & 69 $\pm$ 8 & 69 $\pm$ 11  \\
\enddata
\tablenotetext{a}{The fraction of elliptical and S0 galaxies with
  luminosities $M_V < M^{\star}_{V} + 1$  in the density range.}  

\label{mdrs_lum}
\end{deluxetable*}

\begin{deluxetable*}{lllll}
\tablecolumns{5}
\tablecaption{Mass-Selected Morphology-Density Relations}
\tablehead{\colhead{Cluster} & \colhead{} &  
  \colhead{ $1580> \Sigma > 500$\tablenotemark{a}} &
  \colhead{ $500 > \Sigma > 167$\tablenotemark{a}} &
  \colhead{ $167> \Sigma > 50$\tablenotemark{a}}  \\
\colhead{} & \colhead{} &  
  \colhead{${\rm galaxies\ Mpc^{-2}}$} &
  \colhead{${\rm galaxies\ Mpc^{-2}}$} &
  \colhead{${\rm galaxies\ Mpc^{-2}}$} \\
\colhead{} & \colhead{$z$} & \colhead{(\%)} &
  \colhead{(\%)} & 
  \colhead{(\%)} \\
}
\startdata 
Coma & 0.023 & 96 $\pm$ 5 & 99 $\pm$ 3 & 80 $\pm$ 8  \\
\clt & 0.328 & 87 $\pm$ 7 & 88 $\pm$ 5 & 79 $\pm$ 8\\
\mst & 0.587 & 77 $\pm$ 10 & 81 $\pm$ 9 & 60 $\pm$ 13 \\
\ms\ \& \cl & 0.83 & 88 $\pm$ 7 & 75 $\pm$ 8 & 82 $\pm$ 10 \\
\enddata
\tablenotetext{a}{The fraction of elliptical and S0 galaxies with
  masses $> 10^{10.6} M_{\sun}$  in the density range.}

\label{mdrs_mass}
\end{deluxetable*}

\subsubsection{Evolution of the Mass-Selected Early-type Fraction}

Again, as for the luminosity-selected sample, we split the sample into
high- and low-density bins to establish the trend with redshift in the
early-type galaxy fraction.  We measured the mass-selected early-type
fractions in the same two density bins as Figure \ref{fraclz}.  This
is shown in Figure \ref{fracz}.  Clearly the trend with redshift seen
in Figure \ref{fraclz} is much weaker in Figure \ref{fracz}. At high
densities, we find that the early-type galaxy fraction decreases by
-0.13 $\pm$ 0.09 $z$ while for the low density sample we find -0.14
$\pm$ 0.10 $z$. These slopes are shown as dashed lines in Figure
\ref{fracz}.  Neither represents evolution at the 95\% confidence
limit. We note here that the errors are the same size as those given
earlier for the luminosity-selected samples, but the slopes are half
the size.  We find mean early-type fractions of \ldf\ $\pm$ 3\% at $
500\ {\rm galaxies\ Mpc^{-2}} > \Sigma > 50\ {\rm galaxies\ Mpc^{-2}}$ and \hdf\
$\pm$ 4\% at $\Sigma\ > 500 \ {\rm galaxies\ Mpc^{-2}}$. This reaffirms the
lack of evolution seen in Figure \ref{mdr_mass} and Figure
\ref{mdr_massd}.  We show the mean value of the early-type fraction as
a solid line.  The early-type fractions for the mass-selected sample
in Figure \ref{fracz} are $\sim$20\% higher than the $z> 0.3$ results
from the literature shown in Figure \ref{fraclz}, and $\sim$10-20\%
higher than our own results from the luminosity-selected samples.
These results can be seen in fifth and sixth columns of Table
\ref{results} as well.

We measured the slopes for the mass-selected local galaxy densities in
two density bins as well.  We simply use $250\ > \Sigma > 25\ {\rm
galaxies\ Mpc^{-2}}$ and $\Sigma\ > 250 \ {\rm galaxies\ Mpc^{-2}}$ as the
typical mass-selected density is half that of the luminosity-selected
density.  We find slopes of -0.13 $\pm$ 0.12 $z$ at high densities and
-0.16 $\pm$ 0.12 $z$ at low densities. The fraction of early-type
galaxies in the high density is 89 $\pm$ 4\% while we find 82 $\pm$
3\% in the low-density bin.  These are the same within the errors as
was found for the early-type fractions in the luminosity-selected
density bins.

The lack of a trend in the mass-selected sample points to galaxies of
mass less than  $10^{10.6} M_{\sun}$ as being the likely contributors
to the evolution at high redshift seen in the luminosity-selected
samples, as shown in Figure \ref{mdr_lum} and Figure \ref{fraclz} This
would be consistent with expectations from the increasing global
star formation rate at higher redshift.  The resulting substantial
changes in their M/L would lead to low-mass galaxies brightening into
the luminosity-selected samples.

\begin{deluxetable*}{llllll}
\tablecolumns{6}
\tablecaption{Early-type Galaxy Fractions in Two Density Bins}
\tablehead{\colhead{Cluster} & \colhead{} &
  \colhead{$\Sigma > 500$\tablenotemark{a}} &
  \colhead{$500 > \Sigma > 50$\tablenotemark{a}} & 
  \colhead{$\Sigma > 500$\tablenotemark{b}}  &
  \colhead{$500 > \Sigma > 50$\tablenotemark{b}} \\
\colhead{} & \colhead{} & \colhead{${\rm galaxies\ Mpc^{-2}}$} &
  \colhead{${\rm galaxies\ Mpc^{-2}}$} & \colhead{${\rm galaxies\ Mpc^{-2}}$}  &
  \colhead{${\rm galaxies\ Mpc^{-2}}$} \\
\colhead{} & \colhead{$z$} & \colhead{(\%)} &
  \colhead{(\%)} & \colhead{(\%)}  &
  \colhead{(\%)} \\
}
\startdata
Coma & 0.023 & 96 $\pm$ 4\% & 86 $\pm$ 4\% & 96 $\pm$ 4\% & 86 $\pm$
4\%  \\
\clt & 0.328 & 86 $\pm$ 8\% & 80 $\pm$ 5\% & 89 $\pm$ 6\% & 85 $\pm$
4\% \\
\mst & 0.587 & 74 $\pm$ 12\% & 57 $\pm$ 8\% & 77 $\pm$ 11\% & 74 $\pm$
7\% \\
\ms\ \& \cl & 0.83 & 83 $\pm$ 7\% & 69 $\pm$ 8\% & 89 $\pm$ 7\% & 77
$\pm$ 7

\enddata
\tablenotetext{a}{The fraction of elliptical and S0 galaxies with
  luminosities $M_V < M^{\star}_{V} + 1$ in the density range.}  
\tablenotetext{b}{The fraction of elliptical and S0 galaxies with
  masses $> 10^{10.6} M_{\sun}$ in the density range.}

\label{results}
\end{deluxetable*}

\subsubsection{Is Coma Unusual?}

A minor but interesting issue arises regarding Coma and its somewhat
higher early-type fraction.  In Figures \ref{mdr_mass}, \ref{fracz}
and \ref{mdr_massd}, Coma appears to have a larger number of
early-type galaxies at the highest densities.  We find that Coma has
an early-type fraction 9\% $\pm$ 6\% higher than the average
early-type fraction in the highest-density bin in Figure \ref{fracz}.
The apparent higher fraction of early-type galaxiess and the lack of
any late-type galaxies at very high densities for Coma could have
several causes.  First, the number of spiral and late-type galaxies in
the other clusters could be overestimated.  \citet{diaferio2001}
provides a cautionary note, stating that in some simulations,
star-forming field galaxies could be projected into a cluster even for
redshift-selected samples.  Using the data from GOODS
\citep{vanderwel2007a}, we estimate that 2$\pm$2 late-type galaxies
above our mass threshold per cluster would fall within the redshift
windows listed in Table \ref{summary}.  This would change the
early-type fraction by 2\% (for \ms\ or \cl) 10\% (for \mst),
depending on the richness of the cluster and the density bin in
question.  This would raise the higher redshift cluster early-type
fractions, to an average of 89\%, closer to the Coma value.  A second,
more interesting possibility, is that there is evolution in the number
of spirals in the cores of clusters from $z\simeq 0.3$ to $z\simeq 0$.
This would be mild evolution, 9 $\pm$ 6\% at the highest densities we
sample, as compared with the changes of 25\%-30\% seen in previous
luminosity-selected samples, such as \citet{dressler97},
\citet{postman2005} or \citet{desai2007}.  Third, Coma could be on the
high side of the distribution of early-type fraction among
low-redshift clusters, as indicated in Figure \ref{mdr_lum}.  The
significance of contributions from these effects could be evaluated
better with larger numbers of clusters and larger mass-selected galaxy
samples.

\begin{figure}[htbp]
\begin{center}
\includegraphics[width=3.4in]{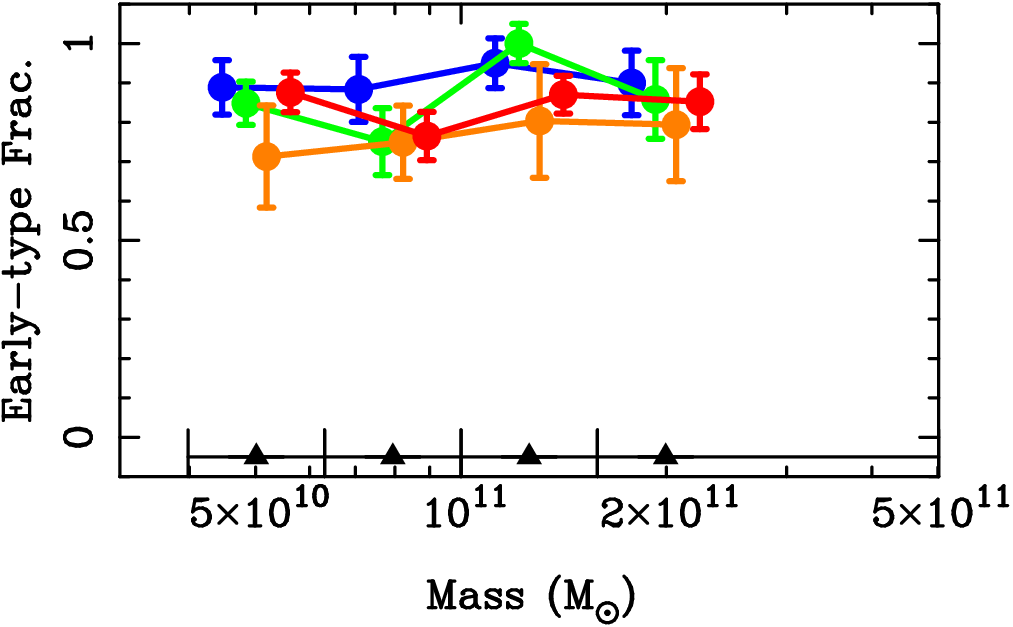}
\end{center}

\caption[f8.eps]{ Morphology-mass relation for all five clusters.
  Each mass bin is 0.2 dex in width, starting at $10^{10.6}\
  M_{\sun}$.  The data points represent the fraction of early-type
  galaxies with Coma (blue circles), \clt\ (green circles), \mst\
  (orange circles) and both \ms\ and \clt\ (red circles).  The data
  points are offset in mass about the mean in the bin for clarity. The
  triangles show the bin centers. The last bin contains all galaxies
  with masses $>10^{11.2}\ M_{\sun}$. We examined this distribution to
  see if the early-type galaxy fraction did change at, for example,
  the lower end of our mass-distribution.  We find no statistically
  significant slope in the early-type galaxy fraction with mass, and
  the average fractions agree with those we find in different density
  bins, see Table \ref{mass_frac}.  The constant early-type fraction
  we see in Fig. \ref{mdr_mass}, Fig. \ref{mdr_massd} and
  Fig. \ref{fracz} comes  from contributions over the whole of the mass function, and is not caused by just the peak of the mass function is dominated by
  early-type galaxies. The typical mass of a late-type galaxy
  contributing to Figures \ref{mdr_lum} and \ref{fraclz} must be below
  our mass limit of $10^{10.6}\ M_{\sun}$.   }
\label{fracz_m}
\end{figure}

\begin{deluxetable*}{llllll}
\tablecolumns{6}
\tablecaption{Early-type Galaxy Fractions in Mass Bins}
\tablehead{\colhead{} & \colhead{} & 
  \colhead{$M >10^{11.2}$ \tablenotemark{a}} &
  \colhead{$10^{11.1} M_{\sun}$ \tablenotemark{a}} &
  \colhead{$10^{10.9} M_{\sun}$ \tablenotemark{a}}  &
  \colhead{$10^{10.7} M_{\sun}$ \tablenotemark{a}} \\
\colhead{Cluster} & \colhead{$z$} & \colhead{\%} & \colhead{\%} &
\colhead{\%} & \colhead{\%}  \\
}
\startdata
Coma & 0.023 & 89 $\pm$ 6 & 96 $\pm$ 4 & 88 $\pm$ 7 & 89 $\pm$
5  \\
\clt & 0.328 & 86 $\pm$ 10 & 100 $\pm$ 5 & 75 $\pm$ 8 & 85 $\pm$
6 \\
\mst & 0.587 & 79 $\pm$ 14 & 80 $\pm$ 14 & 75$\pm$ 9 & 71 $\pm$
13 \\
\ms\ & 0.83 & 85 $\pm$ 9 & 87 $\pm$ 8 & 76 $\pm$ 9 & 87
$\pm$ 8 \\
and \cl & &  & & & \\
\enddata
\tablenotetext{a}{The fraction of elliptical and S0 galaxies in the listed
  mass range at all local galaxy densities; the binwidth is 0.2 dex
  except for the highest mass bin, which includes all galaxies above
  the mass limit.}  
\label{mass_frac}
\end{deluxetable*}

\subsubsection{The Mass Dependence of the Early-type Fraction}

As a further check on the lack of evolution, we investigated whether, in
some particular mass range, we could see evolution in the early-type
fraction. We subdivided our sample into different mass bins, and
derived the early-type fraction in Figure \ref{fracz_m}.  We found no
significant change of the early-type fraction with mass.  Linear fits
to the data in Table \ref{mass_frac}, the fraction of early-type
galaxies in each mass bin, yield slopes of 0.09 $\pm$ 0.12 for Coma,
0.23 $\pm$ 0.15 for \clt, 0.15 $\pm$ 0.30 for \mst, and 0.01 $\pm$
0.13 for the combined sample of \ms\ and \cl.  For all five clusters,
we find no statistically-significant variation from a slope of zero.
The offsets in the mean with increasing cluster redshift can also be
seen to be very small (consistent with the lack of evolution seen in
Figure \ref{fracz} for the mass-selected sample).  In general, the
lack of a change in the early-type fraction seen in Figures
\ref{mdr_mass}, \ref{mdr_massd} and \ref{fracz} appears to be true for
the whole of the sample, not just a subset of galaxies in a certain
mass range.

We also tested whether changing the mass limit at lower redshifts made
any difference.  This arises from a potential concern that we may have
introduced a systematic bias in our results by keeping a constant mass
limit for all redshifts, given the expectation of some systematic
increase in the mean mass of galaxies (from merging and/or star
formation).  We increased the limit for Coma by a factor 2$\times$ to
correspond to a significant mass buildup between $z\simeq 0.8$ and
$z\simeq 0$. This would accommodate any likely mean mass buildup
\citep[e.g.  for the field ][ finds that $\sim$ 50\%-70\% of today's
galaxies with masses $>10^{10.5} M_{\sun}$ may have undergone mergers
since $z\sim 0.7$]{bell2006}.  The changes in the early-type fraction
were insignificant, as would be expected from Figure \ref{fracz_m}.
This is discussed further in \S 4.1.

\section{Results and Discussion}

\subsection{Mass-Selected and Luminosity-Selected Samples}

As discussed in the \S 1, we used mass-selected samples because we
expect only modest and systematic evolution in galaxy masses with
time, unlike luminosity-selected samples where galaxies can enter and
leave the sample as bursts of star formation dramatically increase the
luminosity for a period.   A mass-selection provides a more robust
means of identifying likely progenitors of $z=0$ cluster galaxies in
higher redshift clusters.  However, even the mass-selected sample of
higher redshift cluster galaxies will not contain all the progenitors
for two reasons.  First, galaxies will continue to grow in mass,
either through star formation or by merging.  Mass growth will bring
objects from below our mass limit at high redshift into the sample at
low redshift.  Second, the early-type fraction is lower in the
surrounding field \citep{noeske2007a, vanderwel2007a}.  Field galaxies
will continue to infall into clusters and build up the cluster
population.  Thus, even if there is no evolution from $z=0.83$ to the
current epoch in the fraction of early-type galaxies in mass-selected
samples, {\it we do not expect that the total mass in early-type
galaxies remains constant.} The total mass in early-type galaxies in the
cluster environment is likely to increase \citep[as has been suggested
for the field by ][for example]{bell2004, faber2005, brown2007} -- it
just has to increase through processes which maintain a similar high
fraction of early-type galaxies at masses $> 10^{10.6} M_{\sun}$ at
any time over the last 7 Gyr.  

It is interesting to explore the types of galaxies in our samples.
Our mass-selected samples of cluster galaxies show little or no change
in the early-type galaxy fraction with redshift for masses $>
10^{10.6} M_{\sun}$, i.e., for masses above $\sim$25\% of the $z=0$
characteristic mass, or ${\cal M_{\star}}$.  This mass limit
encompasses the majority ($\simeq$70\%) of the stellar mass in
galaxies in these clusters. When we use the same parent sample but
select galaxies by luminosity, we recover the MDR and
the evolution seen by other authors.  We illustrate this selection for
the combined sample of \ms\ and \cl\ in Figure \ref{mass_dens} where
we show the distribution of masses and galaxy morphological types over
a range of densities for galaxies in a luminosity-selected sample (all
galaxies have $M_V < M^{\star}_{V} + 1$).  At masses lower than
$10^{10.6}\ M_{\sun}$, this sample of galaxies becomes dominated by
late-type galaxies.  These lower-mass, late-type galaxies do not fall
in our primary sample - because we are incomplete at these low masses
- but do fall into the luminosity-selected sample, and so must be the
primary contributers to evolution in luminosity-selected samples.

\subsection{Merging and Mass Assembly}

Merging may play an important role in the formation of massive, red
galaxies between $z\simeq 1$ and today.  \citet{vandokkum1999} found
that the merger rate in \ms\ was significantly higher than in $z\simeq
0$ clusters, and that the large majority of these are ``dry'' mergers.
\citet{bell2006} finds that, in the field, as many as 50\%-70\% of
today's galaxies with masses $>10^{10.5} M_{\sun}$ could have
undergone mergers since $z\simeq 0.7$, although \citet{masjedi2006}
finds little evidence of mergers among massive, non-star forming
galaxies at $z<0.36$.  

Merging raises an interesting issue.  Could a significant fraction of
the $z\simeq 0$ cluster population lie below our mass limit at high
redshift? As the following arguments show, this appears to be
unlikely.   One estimate of the merger rate is the number of close
pairs of galaxies.  \citet{tran2005} and \citet{postman2006} searched
for close pairs of galaxies in \ms\ and \cl. Close pairs are defined
in \citet{postman2006} as galaxies within $30\ h^{-1}\ {\rm kpc}$, but
more than $200\ h^{-1}\ {\rm kpc}$ from the brightest cluster galaxy.
Our sample includes a subset of those pairs from \citet{postman2006},
as we required both galaxies to be members of \cl, and the galaxies to
be separated by less than 300 ${\rm km\ s^{-1}}$ in the rest-frame of
the cluster, the same criterion of the whole sample discussed in
\citet{tran2005}.  These two samples of potential mergers, 14 in \ms\
and 9 in \cl , resulted in 23 galaxies in pairs or triplets likely to
merge in the next Gyr, or 12\% of our $M_{V} < M^{\star}_V + 1.$
sample.  The fraction of merger candidates in the rest of our sample
is small, $<$1\% \citep{vandokkum1999}, pointing to the possibility
that cluster galaxy mergers were more prevalent either at higher
redshift or in certain kinds of clusters.  Nonetheless, 12\% could
represent the rate of merging in $z \simeq 0.8$ clusters.

\begin{figure}[htbp]
\begin{center}
\includegraphics[width=3.4in]{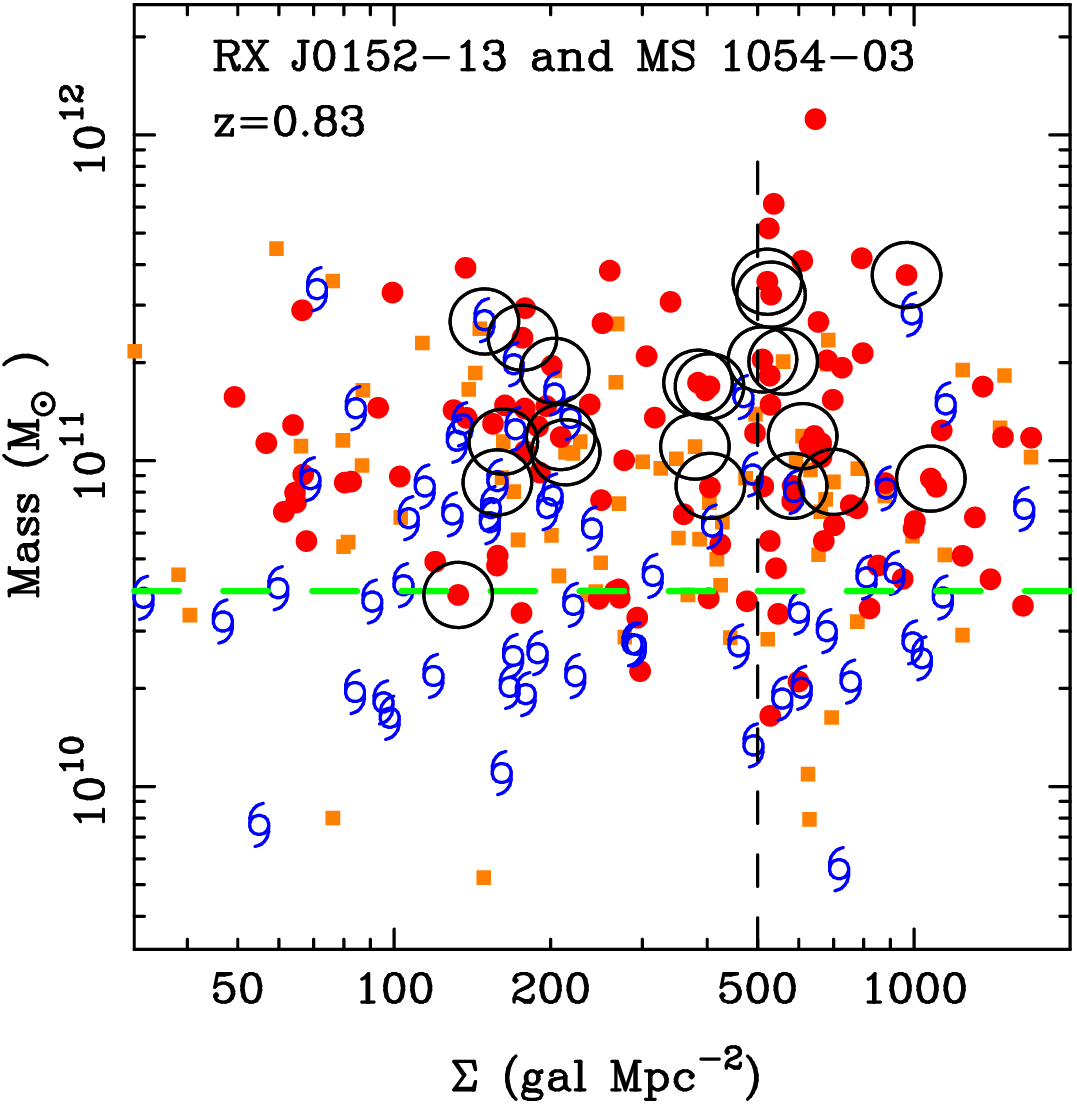}
\end{center}

\caption[f9.eps]{ Mass-density relation for \ms\ and \cl\ at $z=0.83$
  for a magnitude-limited sample.  The red circles are elliptical
  galaxies, the orange squares are S0's and the blue spirals are
  late-type galaxies.  Each galaxy has $M_V < M^{\star}_{V} + 1$, the
  criterion we use for luminosity-selected samples.  The horizontal
  green dashed line shows our mass selection limit while the vertical
  black dashed line shows the division between the two density bins in
  Figs.\ \ref{fraclz} and \ref{fracz}.  Black circles highlight red
  galaxies that are potential mergers in \ms\ \citep{tran2005} and in
  \cl\ \citep{postman2006}.  The mean mass of a late-type galaxy, in
  our luminosity-selected sample, is at the green line, or $10^{10.6
    \pm 0.4} M_{\sun}$, one quarter of the mean mass, $10^{11.2\pm
    0.4} M_{\sun}$, of the early-type galaxies in our
  luminosity-selected sample.  A significant fraction of the late-type
  star-forming galaxies are below our mass limit, yet they would
  be included in the luminosity-selected, MDR for these $z=0.83$
  clusters. At lower redshifts, as star formation ends and they fade
  to lower luminosities, they will drop out of the luminosity-selected
  MDR.  The result would be an apparent increase in the early-type
  fraction in the luminosity-selected samples. }

\label{mass_dens}
\end{figure}

We show the 23 members of these close pairs of galaxies as black
open circles in Figure \ref{mass_dens}.  The average mass of a galaxy in
these pairs, for galaxies above our luminosity limit, is $10^{11.2 \pm
  0.3} M_{\sun}$. The mean early-type galaxy in the whole sample is
the same mass, $10^{11.2 \pm 0.4} M_{\sun}$.  Therefore, the sample of
close pairs in \citet{tran2005} and \citet{postman2006} appear to have
masses typical of ${\cal M_{\star}}$ at $z\sim 0$ cluster early-type
galaxies.  It is possible that the progenitors of some low-mass (close
to our limit) $z\simeq 0$ early-type galaxies have a mass at $z\simeq
0.8$ such that they would fall below our $>10^{10.6} M_{\sun}$ limit.
The close pairs we observe suggest this possibility, but most (19
of 23) of the galaxies observed have masses significantly more than a
factor of 2 above our mass threshold.  Second, only two are spirals,
pointing to a prevalence of ``dry'' merging.  Assuming no new
galaxies, ``dry'' merging will decrease the number of early-type galaxies
above our mass-limit.  However, galaxies below our mass limit could
also merge.  The overall impact from merging on the early-type galaxy
population depends on a number of factors, and requires up-to-date
galaxy formation models to disentangle.  However, if the close pairs
in our sample are representative of the typical mergers between
massive galaxies, the progenitors of the $z\simeq 0$ galaxies near our
mass limit may not have built up through massive mergers and merging
will not have dramatically changed the early-type galaxy fraction.

As mentioned in \S 3 we carried out a test to see if setting the mass
limit in Coma at a higher value, $>10^{10.9} M_{\sun}$, made any
difference (compared to $>10^{10.6} M_{\sun}$ for the $z = 0.83$
clusters).  This would correspond to a large $\simeq 2$times mass
buildup from $z\simeq 0.8$ to $\simeq 0$ \citep[c.f.][results
mentioned above]{bell2006}.  Not unexpectedly, the results are
unchanged (see Figure \ref{fracz_m}).  The early-type galaxy fraction
increases, as would be expected, but very modestly by 4\% at high
densities and by only 1\% at low densities, well within the errors.
Our results will not be changed by choosing time dependent mass
limits, at least for levels that correspond to any likely evolution in
the mean mass buildup in clusters.

\subsection{Field Galaxy Infall}

Infall of late-type galaxies from the low-density region surrounding
the cluster is expected to play a continuing role in the build-up of
early-type cluster galaxies through their transformation to elliptical
and S0 galaxies (and thus leads to a build-up in the total mass in
early-type galaxies with time).  Since the early-type fraction in the
field above our mass limit $10^{10.6} M_{\sun}$ is 30\% lower than in
the cluster \citep[$\simeq$50\%][]{noeske2007a, vanderwel2007a}, the
infalling spiral and late-type galaxies must transform to early-type
galaxies if the infall rate adds significant mass to the cluster (even
if the mechanisms of the transformation are unclear at this time).
Otherwise the fraction of early-type galaxies would decrease with
time, which is inconsistent with what is seen in either the mass- or
luminosity-selected samples.  One caveat on this argument is that the
low-density regions around rich clusters are likely to be unlike the
``true'' field \citep[see e.g.][]{diaferio2001}, and so studies of the
cluster environment within $\sim$ 10 Mpc (comoving) will be needed for
quantitative discussions of the role of infall.

We expect clusters of galaxies to be continually growing in mass by
the infall of galaxies and groups of galaxies from larger radii.  Yet
we find no change in the morphological mix above our mass limit.  No
change in the fraction of early-type galaxies does not preclude
significant mass growth, however.  A simple argument based on the
rough constancy of the observed $\sim$85\% early-type fraction allows
us to estimate how the mass in early-type cluster members could grow.
First, we assume that the infall rate is always high enough to keep
the fraction of late-type galaxies $\simeq$15\%.  Second, we assume
that galaxies take $\simeq$ 1.5 Gyr to transform from a late-type
spiral into an early-type galaxy, the assumption from
\citet{pvd_mf2001}.  Between $z=0.83$ and $=0$, there are 7 Gyr,
slightly more than 4 times the transformation timescale.  Every 1.5
Gyr, the mass in early-type galaxies in this scenario increases by
18\%, as 15\% of the total population of galaxies transforms into same
type as the remaining 85\% of the population.  These assumptions lead
to a doubling of the $z\simeq 0.8$ mass by $z=0$. In other words,
$\simeq$50\% of the $z\simeq 0$ early-type galaxy population enters
the cluster as late-type galaxies between $z\simeq 0.8$ and $z\simeq
0$.  Interestingly, this rate of increase in the mass density of
ellipticals and S0 galaxies is roughly comparable to that typically
measured in field surveys \citep[see][for example]{bell2004,
  faber2005, brown2007}.  The time-scale of transformation and the
infall rate are the largest uncertainties. A longer transformation
time-scale would in turn lower the amount of inferred evolution among
early-type galaxies above our mass-threshold, and this calculation
neglects that many infalling massive galaxies would already be
elliptical and S0 galaxies.

\subsection{The Role of Low -Mass Galaxies -- Transformation at Low Masses}

Our mass-selected sample reaches down to masses $10^{10.6} M_{\sun}$,
and the lack of evolution of the early-type fraction in that sample
leads us to infer that changes at lower masses are playing a role in
the evolution seen in luminosity-selected samples, a possibility 
discusssed in \citet{depropris2003}.  No comprehensive
studies have been carried out that are complete from high to very low
galaxy masses.
 
However, there is some evidence of evolution at low masses
(transformation from late-type to early-type) but the results have not
been consistent. For example, a number of authors have argued for
evolution in the faint, red cluster population \citep{delucia2004b,
  goto2005, tanaka2005, delucia2007b, stott2007}.  This evolution is
observed as an increase in the ratio of faint -- often called
``dwarf'' -- red-sequence galaxies to the ``giant'' galaxies with
luminosities of ${ L^{\star}}$.  In \citet{holden2006}, we
speculated that these additional ``dwarf'' galaxies would become
low-mass S0 galaxies in $z=0$ clusters.  The samples of
\citet{delucia2007b} and \citet{stott2007} are selected to contain
galaxies with masses below $10^{10.6} M_{\sun}$, so there is no
conflict between the observed evolution in that sample for galaxies
with masses of $\simeq 0.25 {\cal M^{\star}}$ and the lack of
evolution in our mass-selected sample containing $> 0.25 {\cal
  M^{\star}}$.  However, the reported results are not unanimous.
\citet{andreon2006} finds no change in the population of faint
galaxies in the cluster \ms.  \citet{tanaka2005} and
\citet{tanaka2007} finds a similar result, i.e., no change in the
luminosity function for the red-sequence in \cl\ and RDCS~1252.9-2927
at $z=1.24$.  Interestingly, \citet{tanaka2005,tanaka2007} find a
truncation in the red sequence for the lower mass groups of galaxies
outside of the cluster cores.  \citet{tanaka2005} and
\citet{andreon2006} both imply that one might find no evolution in the
early-type galaxy fraction at even lower masses for the clusters in
our sample, a result that can be verified with deeper spectroscopic
samples of mass-selected galaxies.

The inconsistencies in these results suggest that robust,
mass-selected samples that are complete to substantially below
$>10^{10.6} M_{\sun}$ are needed to establish the nature of the
evolution of the star-forming galaxies in the cluster environment.

\section{Summary}

We selected a sample of clusters spanning a range in redshift from
$z=0.023$ to $z=0.83$.  Each cluster we included has approximately the
same X-ray luminosity and velocity dispersion to mitigate the observed
correlations between cluster mass estimates and the fraction of
elliptical and S0 galaxies seen in \citet{postman2005} and
\citet{desai2007}.  Using high-quality photometry, largely from {\em
HST}, we computed rest-frame optical colors for the spectroscopically
confirmed members and derived mass-to-light ratios($M/L$).  From 
$M/L$, we determined stellar masses of the \ntot\ galaxies in our
sample.  We find that these stellar masses agree well with those from
dynamical measurements, with a scatter ranging from 0.11 dex at
$z\simeq 0$ to 0.20 dex at our highest redshifts.

All of the galaxies in our sample have morphological classifications.
While the classifications come from a variety of sources
\citep{dressler1980b, fabricant2000, tran2003, tran2004, postman2005},
all of the $z>0.1$ catalogs were constructed to be consistent with the
classification scheme of \citet{dressler1980b}.  We used these
catalogs to derive the morphology-density relations (MDRs) in the five
clusters, and to establish the evolution in that relation for
luminosity-selected samples (for luminosities $<M^{\star}_{V} + 1$.)
As shown in Figures \ref{mdr_lum} and \ref{fraclz}, we found good
agreement for both the $z=0$ MDR from \citet{dressler1980a} and
\citet{postman1984}, as well as for the evolution seen in previous
work \citep{dressler97, smith2005, postman2005}, for the \nlum\
galaxies brighter than $<M^{\star}_{V} + 1$ in our sample.

We contrast the results for our luminosity-selected samples with a
similar large completeness-corrected mass-selected samples of \nmass\
galaxies with masses $>10^{10.6} M_{\sun}$ ($0.25\ {\cal M_{\star}}$
at $z=0$) in each of the clusters. At this mass limit, the majority
($\simeq$70\%) of the stellar mass in galaxies in the clusters is in
our mass-selected sample. 

Very interesting differences are seen between the mass- and
luminosity-selected MDRs.  First, the MDRs derived for the individual
clusters from the mass-selected galaxy samples typically have larger
early-type galaxy fractions, by 10\%-20\%, and show marginal evidence
of a lower slope with density.  Second, and particularly striking, is
that essentially no evolution is seen in the mass-selected early-type
galaxy fraction with redshift.  The change of the slope is
substantially less than for the luminosity-selected sample, and is
consistent with no change with redshift.  We find mean early-type
fractions, \ldf\ $\pm$ 3\% at $ 500\ {\rm galaxies\ Mpc^{-2}}> \Sigma
> 50\ {\rm galaxies\ Mpc^{-2}}$ and \hdf\ $\pm$ 4\% at $\Sigma\ > 500
\ {\rm galaxies\ Mpc^{-2}}$.  This indicates that the evolution in the
early-type fraction observed in luminosity-selected samples of
galaxies arises predominately from the inclusion of lower mass,
$<10^{10.6} M_{\sun}$, late-type galaxies at higher redshifts.

Despite finding no evolution in the mass-selected early-type fraction
with redshift, we recognize that the cluster galaxy formation process
is not finished at $z=0.83$.  The overall mass in early-type galaxies
is expected to increase, but this increase must happen so as to leave
the early-type galaxy fraction largely unchanged.  The early-type
fraction in the field above the same mass limit is very significantly
lower \citep[ 48\% $\pm$ 7\% at $z=0.8$][]{vanderwel2007a}.  Infall of
field galaxies, with a larger fraction of spiral and late-type
galaxies, into clusters is expected and must be accompanied by
transformation to early-type galaxies to ensure that the fraction of
early-types remains constant.

Mass-selected samples to low masses, along with detailed morphologies,
are realizable and would be of great value for characterizing and
constraining galaxy evolution over the last 7-8 Gyr out to $z\sim 1$.
Identifying these physical processes and the nature of the transition
from the field early-type fraction is the next step -- for a range of
cluster masses \citep[see e.g.,][]{postman2005, desai2007}.  Of
particular value will be the outer regions, where future cluster
members will be transformed during their infall into the cluster by
$z\simeq 0$. Significant insights into the build-up of galaxies in
dense environments over the last 7-8 Gyr would result from increasing
the sample of clusters with mass-selected galaxies that have
redshifts, morphologies and excellent photometry.  These data would be
of particular value for developing insight into the formation of S0
galaxies which, show strong evolution in luminosity-selected samples
from $z\simeq 1$ to $\simeq 0$.

In summary, we have derived luminosities, masses and morphologies for
a large and complete sample of cluster galaxies ranging from $z=0.023$
to $0.83$.  The luminosity-selected galaxy samples show, in all
clusters, a clear MDR.  Furthermore, the
early-type galaxy fraction evolves substantially with redshift, as
found by others.  When we construct mass-selected samples, the
MDR  are offset towards larger early-type
fractions and are marginally weaker.  More dramatically, the evolution
in the early-type galaxy fraction is very weak, being consistent with
no evolution over the 7 Gyr period.  As these two samples are drawn
from the same parent sample of five massive clusters of galaxies, we
conclude that the galaxies that lead to the evolution of the
luminosity-selected MDR must be galaxies less
than our mass-limit, namely those with masses $<10^{10.6} M_{\sun}$.

\vspace{.3cm}

We would like to thank Erica Ellingson, Sandy Faber, Justin Harker,
Susan Kassin, David Koo, Anne Metevier, Kai Noeske, S. Adam Stanford,
and Pieter van Dokkum for useful discussions.  We would like to thank
the anonymous referee for helping us substantially improve this paper.
The Advanced Camera for Surveys was developed under NASA contract
NAS5-32865, and this research was supported by NASA grant NAG5-7697.
We are grateful to K.~Anderson, J.~McCann, S.~Busching, A.~Framarini,
S.~Barkhouser, and T.~Allen for their invaluable contributions to the
ACS project at Johns Hopkins Univeristy.  Some of the data presented
herein were obtained at the W.M. Keck Observatory, which is operated
as a scientific partnership among the California Institute of
Technology, the University of California and the National Aeronautics
and Space Administration. The Keck Observatory was made possible by the
generous financial support of the W.M. Keck Foundation.  The authors
wish to recognize and acknowledge the very significant cultural role
and reverence that the summit of Mauna Kea has always had within the
indigenous Hawaiian community.  We are most fortunate to have the
opportunity to conduct observations from this mountain.  This research
has made use of the NASA/IPAC Extragalactic Database (NED), which is
operated by the Jet Propulsion Laboratory, California Institute of
Technology, under contract with the National Aeronautics and Space
Administration.

\appendix

\section{FIRES Data for \ms}
\label{fires}

We used additional photometry for \ms\ from the Faint Infrared
Extragalactic Survey (FIRES), discussed in
\citet{forster2006}.  The FIRES team used FORS1 on the Very
Large Telescope (VLT) to obtain the {\em UBV} data and ISAAC to obtain
the $J_s H K_s$ data.  The ISAAC imaging data consists of four
overlapping pointings with 77 hr of total exposure covering a 5\farcm
5 by 5\farcm 3 field of view.  

The original FIRES catalog was created using a combination of the VLT
data and older WFPC2 imaging in F606W and F814W.  We elected to
replace the WFPC2 data with our ACS imaging, as our data should be
of higher quality and include an additional filter, $z_{850}$.  To
obtain accurate colors, we smoothed the ACS imaging, along with the
rest of the VLT imaging, to the same effective seeing as the worst of
the ground based imaging.  We then re-binned the ACS data to the same
scale, 0\farcs 0996 pixel$^{-1}$, and orientation as the VLT $K_s$ image.
We used the un-smoothed $K_s$ image as a detection image.  All
magnitudes for the colors were measured in fixed apertures on the
smoothed data.  The apertures ranged in size from 1\farcs 0 to 2\farcs
4 in size in increments of 0\farcs 2.  We selected the aperture in
which the signal-to-noise ratio for the color measurements was maximized.
We used SExtractor \citep{bertin96} for object detection and
photometry for the FIRES imaging.

As we have re-measured the magnitudes in all the images, we compared
our resulting magnitudes with those as measured by the FIRES team.
For the near infrared data, the scatter was less than 0.01 mag.  For
the $B$ and $V$ data, the scatter was around 0.02 mag, with the $U$
showing 0.04 mag in scatter.  Given that the images are the same, this
scatter came from different choices of parameters used for object
detection parameters, resulting in slightly different object positions
and sizes.  We examined the scatter for those objects labeled by
SExtractor as ``biased'', i.e., objects with close neighbors in the
FIRES catalog.   We found that the scatter increased
(at times it doubled) only for those objects flagged as ``biased''.
Thus, the difference in detection parameters is most likely what
causes the scatter in the photometry between our measurements and
those of the FIRES team.  In addition to the scatter, we found a small
offset between our magnitudes and those from FIRES.  This result came
from the different point spread function smoothing kernels used by us
and by the FIRES team.  This offset is less than 0.01 mag for all
filters.

Our photometric catalog was then merged with the ACS imaging catalog.
We selected galaxies from the FIRES catalog, then matched those to
identifications from the ACS catalog. This catalog contains
morphological types from \citep{postman2005} and the colors from 
\citet{blakeslee2005}.  When two objects from the ACS imaging were
identified as a single object in the FIRES imaging, we removed them.
The final catalog contains 1221 galaxies, with a subset of 482 that
have morphological classifications from \citet{postman2005}.

We found a small $0.018 \pm 0.012$ mag offset between our \z\
magnitudes and those from \citet{blakeslee2005}.  The magnitudes in
\citet{blakeslee2005} have a 0.2 mag aperture correction to estimate
total magnitudes.  The FIRES magnitudes are based on the five Kron
radius aperture magnitudes from SExtractor as measured on the $K_s$
imaging.  The $z_{850}-K_s$ color is then used to compute the total
flux in the \z\ filter.  However, no aperture correction is made, so
this 0.018 mag offset reflects how much light is lost in the $K_s$
imaging.

\section{Mass Estimates Based on Multi-band Photometry}
\label{masses}

For \ms, we estimated stellar masses by fitting spectral energy
distributions (SEDs) to the photometric data from the FIRES data set.
Specifically, we used two sets of BC03 models: simple
stellar populations (SSPs) and models with exponentially decaying star
formation rates.  These were fit to the combination of the FIRES data
and ACS imaging. For both sets of BC03 models, we assumed a Salpeter
initial mass functions.  When using models with exponentially decaying
star formation rates, or $\tau$ models, we assumed timescales
($\tau$) ranging from 100 Myr to 5 Gyr.  For both the SSPs and the
$\tau$ models, we used three different metallicity values, 0.4,
1, and 2.5 $Z_{\sun}$.

Given a SED and the photometry of a galaxy in one of the clusters in
our sample, we redshifted the SED to the observed value.
We computed the fluxes in the observed filters using the filter curves
provided by ESO for FORS and ISAAC, and by the Space Telescope Science
Institute for the ACS \citep{sirianni2005}.  For a given
metallicity and $\tau$, we allowed the range of ages to vary.  For
every model, we computed the normalization $n$ that minimized the
$\chi^2$ for that model. This eliminates the need to fit for the best
model and the best normalization, as the normalization is the one that
automatically minimizes the $\chi^2$ for a given model. The
normalization is \( n = (\sum f_{i}m_{i}/\sigma^{2}_{i})/( \sum
m_{i}^2/\sigma^{2}_{i})\) where $i$ denotes a filter, $f_i$ is
the observed flux, $m_i$ is the model flux and $\sigma_i$ is the
estimated error.  The error on $n$ is \( \sigma_n = (\sum
  \sigma^{2}_{i}/\sum m^2_{i})^{1/2} \).  This normalization provides the
mass in stars required to reproduce the observed flux in the color
aperture.  We used the total magnitudes in the $K_s$ filter to correct
this normalization to a ``total'' stellar mass.  We then compared the
$\chi^2$ for each model using this normalization while varying the
other parameters of interest.

These model fits were also used to derive rest-frame colors.  We
computed the model's $z=0$, or rest-frame, magnitudes in the desired
filters, such as $B$ and $V$.  We then computed the model magnitudes
in the observed filters as part of the $\chi^2$ fitting.  We used the
offset between the best-fitting model's observed and rest-frame
magnitudes to estimate the rest-frame magnitudes from the data for
each galaxy.

For both the SSP and $\tau$ models, we included the effect of dust by
adding an additional foreground screen to our spectral energy
distribution model fits.  The typical change in the resulting stellar
mass was minimal.  For example, the average mass changes by 7\%
between $\tau$ models with dust and those without.  The models
generally fit the data with a minimal amount of dust for most
galaxies, except for the bluer, star-forming objects.  Removing the
filters that sample the rest-frame UV, the $U$ and $B$ in the FIRES
data sample $<$2500\AA at $z=0.83$, reduces this change in the mass to
4\%.  When combined with $\tau$ models, dust removes any
mass-metallicity relation and, as expected, moves the average age to
younger values.  The mode of the age distribution, however, stays the
same.  The modal age of the population is the age of the
``red-sequence'', and those galaxies are all fit with little or no
dust.  The addition of dust adds a tail of younger ages.  Thus dust
models significantly changed the ages and masses only for the bluer
populations, as expected for galaxies with on going star formation,
and do not affect the majority of the cluster population.

\begin{figure}[htbp]
\begin{center}

\includegraphics[width=3.4in]{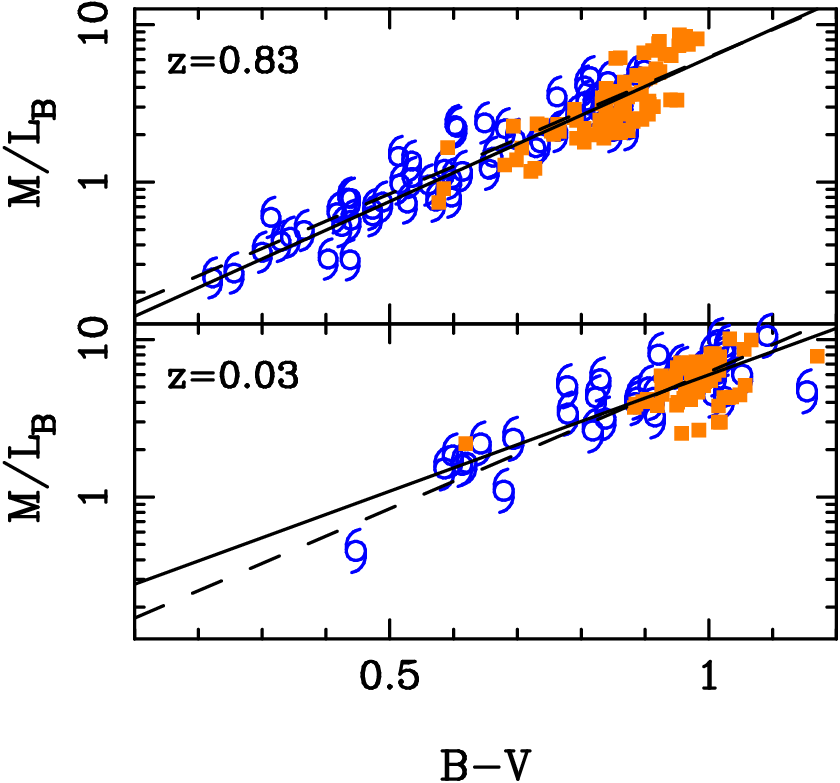}
\end{center}
\caption[f10.eps]{ $M/L_B$ as a function of rest-frame $B-V$ color for
  the \ms\ sample ({\em top}) and the SDSS DR4 sample ({\em bottom}).
  In each panel, blue spirals are late-type galaxies while the orange
  squares are early-type galaxies.  The solid lines represent the best
  fit to the relation between the $B-V$ color and the mass-to-light
  ratio in the $B$ band.  The dashed lines are the relations from
  \citet{bell2003}.  We reproduce the results of \citet{bell2003}, and
  thus used that relation to estimate $M/L_B$ for the whole of our
  cluster sample.  The scatter around the relation is 0.14 dex in the
  high-redshift sample and 0.12 dex for the low-redshift sample, with
  no significant difference in the scatter for early or late-type
  galaxies.}
\label{ml_bmv_both}
\end{figure}

\citet{bell2003} give, in \S 2 of their Appendix A2 (Table 7), a relation
between rest-frame color and $M/L$ in the Johnson $B$, $V$, and
$R$ passbands.  We found, as in \citet{bell2003}, the best-fitting SED
$M/L_B$ correlates strongly with the $B-V$ rest-frame color for the
galaxies in \ms, see Figure \ref{ml_bmv_both}.  This plot is made by
fitting dust-free $\tau$ models to all of the FIRES and ACS filters.
We show our best-fitting relation between $M/L_B$ and the
rest-frame $B-V$ color.  We find that our relations are very similar
to the results from \citet{bell2003}, shown as dashed lines in Figure
\ref{ml_bmv_both}.

We found a scatter of 33\%, or 0.14 dex, in $M/L$ around the
best-fitting relation for \ms.  Early-type galaxies have a slightly
larger scatter of 34\% or 0.15 dex, while late-type galaxies have 30\%
or 0.13 dex.  In our lower redshift sample created using the SDSS DR4,
we find a smaller scatter, 28\% or 0.12 dex, with the same scatter for
both late and early-type galaxies.

\end{document}